\documentclass[
reprint,
nofootinbib,
 amsmath,amssymb,
 aps,
showkeys
]{revtex4-2}

\usepackage{placeins}

\usepackage{graphicx}
\usepackage{dcolumn}
\usepackage{bm}
\usepackage{hyperref}
\usepackage[mathlines]{lineno}

\usepackage{amsthm}

\usepackage{float}
\makeatletter
\let\newfloat\newfloat@ltx
\makeatother

\usepackage{xcolor}
\usepackage{algorithm}
\usepackage{algpseudocode}

\begin{document}

\preprint{APS/123-QED}

\title{Permanent magnet optimization of stellarators with coupling\\ from finite permeability and demagnetization effects}

\author{Armin Ulrich}
\email{Corresponding author: au2171@nyu.edu}
\affiliation{Courant Institute of Mathematical Sciences \\ New York University, New York, NY 10012, USA} 

\author{Mason Haberle}
\affiliation{Courant Institute of Mathematical Sciences \\ New York University, New York, NY 10012, USA}

\author{Alan A. Kaptanoglu}
\affiliation{Courant Institute of Mathematical Sciences \\ New York University, New York, NY 10012, USA}
\begin{abstract}
Permanent magnets provide an attractive path for shaping university-scale stellarator magnetic fields. Previous work has shown that greedy permanent magnet optimization (GPMO) can produce sparse, grid-aligned arrays that match target surfaces with high accuracy under an ideal rigid-remanence model. Here we extend this approach to a greedy permanent magnet optimization with macromagnetic refinement (GPMOmr) by introducing a block-level macromagnetic model that accounts for magnet--magnet and magnet--coil coupling from finite permeability and demagnetizing interactions, and apply it to the published magnet grid from the MUSE stellarator design. Finite-permeability effects produce degree-scale tilts and few-percent magnitude changes in individual magnets and modify the surface-normal field $\mathbf B\cdot\mathbf n$ only at the percent level, yet for a fixed layout they increase the standard squared-flux objective by more than a factor of two. When the same model is embedded in the greedy loop, GPMOmr achieves $f_B$ histories and final errors within a few percent of classical GPMO while producing visibly more nonuniform magnetization patterns. Our formulation provides a fast and practical tool for quantifying and incorporating finite-permeability effects in permanent-magnet stellarator designs, and offers a framework for extending permanent-magnet optimization to higher field strengths and to materials with stronger macromagnetic coupling.
\end{abstract}


\maketitle



\section{Introduction}
Stellarators confine plasma with fully three-dimensional magnetic fields and have long promised steady-state operation without large inductive currents or disruptions~\cite{Boozer2005,Helander2014}. Realizing these advantages at scale historically required intricate, nonplanar superconducting coils whose complexity dominated cost and schedule on flagship devices such as W7-X~\cite{Wolf2017}. Over the last decade, a complementary approach for university-scale experiments has gained growing attention: shaping with permanent magnets (PMs). By allowing simple toroidal-field (TF) coils to supply most of the toroidal flux while PMs provide the fine nonaxisymmetric shaping, one can dramatically simplify engineering without sacrificing optimized physics~\cite{Helander2020PMStell,Landreman2021LLS,Zhu2020PerpPM,Hammond2020GeomPM,Zhu2020,Qian2022,Kaptanoglu2022SparsePM,Hammond2022QUASAR,Lu2021FourierPM,Lu2021TwoStepPM,Lu2022CRPS,Madeira2024TokamakPM}.

A key milestone for this approach is MUSE, the first quasi-axisymmetric stellarator built with a dense array of discrete PM ``towers'' located between a circular TF-coil set and a glass vacuum vessel~\cite{Qian2023}. In MUSE the coil shapes and currents are taken as fixed design inputs; only the permanent-magnet array is optimized, so that the coils are not part of the stellarator-shaping optimization. On MUSE, as for many stellarator designs, the plasma boundary is chosen by a first stage of optimization; in the second stage of optimization, PM distributions are computed (e.g.\ with FAMUS~\cite{Zhu2020,Hammond2020GeomPM}) to minimize the normal-field residual $\mathbf B\cdot\mathbf n$ on the target surface while respecting strong manufacturability constraints (single orientation, finite thickness, ports, and assembly gaps). Recent algorithmic advances in discrete PM optimization, most notably the greedy permanent magnet optimization (GPMO) family, have made such stage-two loops practical at scale by enabling sparse, binary, grid-aligned PM layouts to be computed quickly and at scale, often matching or exceeding results obtained from more elaborate continuous or topology-based optimization pipelines, while remaining practical for engineering use~\cite{Kaptanoglu2023,Kaptanoglu2022SparsePM,Hammond2022QUASAR,Hammond2024CPC}.

Despite this progress, a central question remains insufficiently quantified.
To what extent do \emph{macromagnetic} effects---those arising from finite permeability, demagnetizing interactions, and mutual field--magnetization feedback at the \emph{component} (tower or block) level---perturb the idealized picture of linear superposition that is used during permanent-magnet optimization? This question becomes even more pressing for higher-field and larger PM concepts, where local applied and demagnetizing fields can approach material limits; a common concern is that scaled-up PM stellarators may drive some magnets into substantial (and potentially near-complete) demagnetization, so the ability to evaluate coupled magnetization changes and their impact on plasma metrics is essential. The MUSE design paper~\cite{Qian2023} performed an important first assessment by computing finite-$\mu$ corrections to equilibrium metrics with the tile-based micromagnetic solver MagTense~\cite{Bjoerk2021}. That study showed that anisotropic permeability along the magnetization direction is the dominant effect and can largely be compensated by a small retuning of the toroidal-field current. However, MagTense is tailored to micromagnetic tiles and local demagnetization studies. Running full-device sweeps over $10^4$ to $10^5$ discrete bodies and coupling those fields self-consistently into stellarator metrics (for example, surface $\mathbf B\cdot\mathbf n$ and free-boundary equilibria) is computationally challenging at that scale. Complementary work by Chambliss \emph{et al.} has recently applied gradient and Hessian methods to MUSE and PM4STELL to quantify how idealized perturbations of permanent-magnet dipole moments and positions affect resonant perturbations and magnetic island widths, highlighting the strong sensitivity of PM stellarators to small magnet errors~\cite{Chambliss2025}. While that analysis treats the magnets as independent dipoles and does not include finite-$\mu$ feedback or demagnetizing interactions, it underscores the need to understand how realistic, coupled magnetization changes project onto standard surface and equilibrium metrics. Moreover, while residual macroscopic deviations are often regarded as minor, there has not been a systematic quantification of how much $\mathbf B\cdot\mathbf n$ error arises from effective \emph{tilts} of magnetization vectors compared with \emph{magnitude} changes, nor has the role of coil and PM cross effects been explicitly separated.

This paper aims to address these open questions with three contributions:

\begin{itemize}
\item \textbf{A device-scale macromagnetics solver for coupled block magnetizations.}
\end{itemize}
\noindent
We develop a fast, device-scale \emph{macromagnetics} model that augments binary, grid-aligned PM designs with efficient corrections for finite $\mu$ and demagnetizing interactions at the block/tower level. The model is designed to be composable with stage-two PM-optimization pipelines and scalable to full stellarators (many field periods, ports, and cut planes), while remaining faithful to standard magnetic-material phenomenology~\cite{Jiles2016}.

\begin{itemize}
\item \textbf{A quantitative post-analysis of the published MUSE magnet grid.}
\end{itemize}
\noindent
We perform a full \emph{post-analysis} of the published MUSE PM grid~\cite{Qian2023} to quantify how blockwise \emph{angular} deviations (effective tilts) and \emph{magnitude} changes (effective remanence shifts) separately map into residual $\mathbf B\cdot\mathbf n$ on the target surface and into the squared-flux objective $f_B$. We report the relative contributions of these channels and show that macromagnetic corrections remain percent-level in $\mathbf B\cdot\mathbf n$ but can still change global error measures by order-one, refining earlier qualitative conclusions about their importance.

\begin{itemize}
\item \textbf{GPMO with macromagnetic refinement (GPMOmr) for rapid re-optimization.}
\end{itemize}
\noindent
We implement a \emph{GPMO with macromagnetic refinement (GPMOmr)} loop. Because GPMO is scalable and provides explicit engineering control over discrete, binary magnet choices~\cite{Kaptanoglu2023,Kaptanoglu2022SparsePM,Hammond2022QUASAR,Hammond2024CPC}, this enables rapid re-optimization on large arrays that (i) absorbs macromagnetic deviations directly in the optimization step, (ii) yields layouts with plasma-facing errors comparable to classical GPMO but internally redistributed magnetization, and (iii) quantifies robustness margins for MUSE and explores alternative magnetic materials or layouts for future PM stellarators.

An open-source implementation of the macromagnetic solver and the GPMOmr algorithm is available in the SIMSOPT code base.\footnote{Implementation available at \url{https://github.com/armulrich/simsopt/tree/simsopt_macromag}.}


In short, our goal is twofold: (i) provide a rigorous, quantitative postmortem of MUSE's PM grid with respect to macromagnetic effects, and (ii) deliver a practical, scalable modeling and optimization stack that informs next-generation PM designs and material choices. Together with prior work on topology optimization for PM arrays~\cite{Zhu2020,Zhu2020PerpPM,Hammond2020GeomPM} and greedy discrete optimization~\cite{Kaptanoglu2023,Kaptanoglu2022SparsePM,Hammond2022QUASAR,Hammond2024CPC}, our macromagnetics layer helps bridge idealized design and device reality for permanent-magnet stellarators.



\section{Background}
In this section we briefly review the micromagnetic description of hard magnets, the role of anisotropy and exchange, and the demagnetization tensor, before turning to greedy permanent-magnet optimization.

\subsection{Micromagnetic energy densities and the effective field}
At material length scales small compared to the magnet dimensions but large compared to atomic spacings, the magnetization may be modeled as a continuous field $\mathbf M(\mathbf r)=M_s\,\mathbf m(\mathbf r)$ with $\|\mathbf m\|=1$. The standard micromagnetic Gibbs free-energy density is decomposed as
\begin{equation}
g(\mathbf m,\nabla\mathbf m)
= g_{\mathrm{ex}} + g_{\mathrm{demag}} + g_{\mathrm{ani}} + g_{\mathrm Z},
\label{eq:g_sum}
\end{equation}
with the four principal contributions
\begin{align}
g_{\mathrm{ex}} &= A_{\mathrm{ex}}\|\nabla\mathbf m\|^{2},
\label{eq:g_ex} \\[4pt]
g_{\mathrm{demag}} &= -\frac{\mu_0 M_s}{2}\,\mathbf m\!\cdot\!\mathbf H_d,
\label{eq:g_demag} \\[4pt]
g_{\mathrm{ani}} &= -K_u\,(\mathbf m\!\cdot\!\hat{\mathbf u})^{2},
\label{eq:g_ani} \\[4pt]
g_{\mathrm Z} &= -\mu_0 M_s\,\mathbf m\!\cdot\!\mathbf H_a .
\label{eq:g_zeeman}
\end{align}

Here $A_{\mathrm{ex}}$ is the exchange stiffness, $K_u$ the uniaxial anisotropy constant, $\hat{\mathbf u}$ a unit vector along the local easy axis, $\mathbf H_a$ an externally applied field, and $\mathbf H_d$ the demagnetizing (stray) field determined by the magnetostatic relations $\nabla\!\cdot\!(\mathbf H_d+\mathbf M)=0$ and $\nabla\times \mathbf H_d=\mathbf 0$. The exchange and anisotropy terms originate in quantum-mechanical exchange and spin-orbit coupling, respectively, while $g_{\mathrm{demag}}$ and $g_{\mathrm Z}$ are classical magnetostatic contributions~\cite{Atxitia2017,Bjoerk2021}.


Stationary (or damped) micromagnetic states satisfy Brown's condition $\mathbf m\times\mathbf H_{\mathrm{eff}}=\mathbf 0$ with the effective field obtained by variational differentiation of the total energy $G=\int_\Omega g\,dV$ over the ferromagnetic body $\Omega$:
\begin{equation}
\begin{split}
\mathbf H_{\mathrm{eff}}
  &= -\frac{1}{\mu_0 M_s}\,\frac{\delta G}{\delta \mathbf m} \\[4pt]
  &= \frac{2A_{\mathrm{ex}}}{\mu_0 M_s}\,\nabla^2\mathbf m
     + \mathbf H_d
     + \frac{2K_u}{\mu_0 M_s}(\mathbf m\!\cdot\!\hat{\mathbf u})\,\hat{\mathbf u}
     + \mathbf H_a .
\end{split}
\label{eq:Heff_micro}
\end{equation}
Equation~\eqref{eq:Heff_micro} underlies both dynamic Landau--Lifshitz (Gilbert) solvers and static fixed-point schemes used by contemporary micromagnetic codes, including the tile-based framework of MagTense~\cite{Bjoerk2021}.



\subsection{Easy axis and anisotropy}
In ferromagnetic materials the magnetocrystalline anisotropy introduces preferred directions of magnetization, termed easy axes. Microscopically, this arises from spin--orbit coupling that links the electron spin to the underlying crystal lattice, producing an energy landscape with distinct minima. In the absence of an external field, the magnetic dipoles within each domain preferentially align along one of these equivalent easy axes. The corresponding anisotropy term [Eq.~\eqref{eq:g_ani}] thus acts as a restoring force, penalizing deviations from the easy direction and favoring alignment along it~\cite[Ch.~6]{Jiles2016,Herbst1991R2Fe14B}.



Although the easy axis is not fundamentally immutable, in most technologically relevant rare-earth magnets it may be treated as fixed after fabrication. During industrial processing, Nd--Fe--B powders are exposed to multi-Tesla aligning fields prior to sintering, locking their crystallographic $c$-axes into a common orientation; subsequent magnetization pulses drive the magnetization into a single anisotropy well, leaving a remanent state $\mathbf M_{\mathrm{rem}}=M_{\mathrm{rem}}\hat{\mathbf u}$~\cite{Sagawa1984NdFeB,Herbst1991R2Fe14B,Jiles2016}. Typical anisotropy fields $H_{\rm ani}=2K_u/(\mu_0 M_s)$ for Nd$_2$Fe$_{14}$B lie in the several-Tesla range~\cite{Herbst1991R2Fe14B,Jiles2016}, so ordinary operating fields are too weak to rotate the easy axis, which remains locked to the lattice throughout the lifetime of the magnet. In this sense such materials are ``hard'' magnets, in contrast to ``soft'' magnets where low anisotropy allows continuous reorientation of domains under modest fields~\cite[Chs.~6--8]{Jiles2016}. During stellarator operation the magnetic fields at the permanent-magnet holders in MUSE remain sub-Tesla: the on-axis average field is $\langle|\mathbf B|\rangle \approx 0.15$~T and the local fields at the PM towers are of order $0.2$--$0.3$~T~\cite{Qian2023,Chambliss2025}. These values are well below both the local PM fringe fields and the anisotropy field $H_\text{ani}$, so treating $\hat{\mathbf u}$ as fixed is appropriate for this device-scale study.

\subsection{Landau--Lifshitz equation}
The temporal evolution of the magnetization is governed by the Landau--Lifshitz equation, later generalized to include Gilbert damping. For a normalized magnetization $\mathbf m=\mathbf M/M_s$, the equation reads
\begin{equation}
\frac{\partial \mathbf m}{\partial t}
= -\gamma\,\mathbf m\times \mathbf H_{\mathrm{eff}}
  + \alpha\,\mathbf m\times\frac{\partial \mathbf m}{\partial t},
\label{eq:LLG}
\end{equation}
where $\gamma$ is the gyromagnetic ratio and $\alpha$ a dimensionless damping constant. The precessional term enforces conservation of $|\mathbf m|=1$, while the damping term drives relaxation toward equilibrium states with $\mathbf m\times\mathbf H_{\mathrm{eff}}=0$~\cite{Atxitia2017}. This partial differential equation underlies micromagnetic simulation codes, providing a dynamic description of magnetization switching, relaxation, and domain-wall motion.

\subsection{Demagnetization tensor}
The demagnetizing field $\mathbf H_d$ entering Eq.~\eqref{eq:g_demag} is determined by the long-range dipolar interaction of the magnetization with itself. With a large number of magnets, this relation can be encoded by the demagnetization tensor $N_{ij}(\mathbf r,\mathbf r')$, which connects the local demagnetizing field to the magnetization distribution,
\begin{equation}
H_{d,i}(\mathbf r) = - \int_\Omega N_{ij}(\mathbf r,\mathbf r')\,M_j(\mathbf r')\,dV' .
\end{equation}


 Analytical forms of $N_{ij}$ are known for uniformly magnetized ellipsoids, but for general shapes the tensor must be computed numerically. For rectangular prisms, however, closed-form expressions for the demagnetization tensor field are available: the diagonal terms can be written as sums of $\arctan(\cdot)$ contributions over the eight signed corner combinations, while the off-diagonal terms can be written as $-\frac{1}{4\pi}\ln(\cdot)$ of a product ratio of auxiliary functions. In this work we evaluate each prism--prism interaction block $N_{ij}$ using these analytical expressions for a uniformly magnetized rectangular prism~\cite{Smith2010}. Tile-based frameworks such as MagTense employ such analytical demagnetization kernels evaluated over rectangular cells, which allows accurate resolution of self- and mutual interactions but at significant computational cost when extended to large systems~\cite{Bjoerk2021}. Moreover, $N_{ij}$ is symmetric by construction (reciprocity), $N_{ij}=N_{ji}^{\top}$, which roughly halves the number of unique blocks that must be computed and cached~\cite{Newell1993,Aharoni1998,EngelHerbert2005,Smith2010}. Despite this helpful symmetry, the operator remains dense due to long-range couplings, so a single matrix--vector application costs $O(N^2)$; and once site-dependent anisotropic susceptibilities are incorporated, the assembled equilibrium matrix
\begin{equation}
A_{ij} = \delta_{ij}\mathbf I_3 + \boldsymbol\chi_i\,\underline{\underline N}_{ij}
\end{equation}
is generally nonsymmetric, precluding conjugate gradients and motivating Krylov solvers for nonsymmetric systems (e.g.\ GMRES/biCGStab)~\cite{SaadSchultz1986,Saad2003,Bjoerk2021}.

\subsection{Greedy permanent magnet optimization (GPMO)}

While continuous optimization methods have long been applied to PM field synthesis, they operate in high-dimensional spaces of continuous magnetization variables and must enforce nonlinear constraints to recover discrete, maximum-strength, grid-aligned solutions~\cite{Landreman2021LLS,Zhu2020,Zhu2020PerpPM,Hammond2020GeomPM,Helander2020PMStell,Kaptanoglu2022SparsePM,Hammond2022QUASAR}. Greedy permanent magnet optimization (GPMO) instead treats the discrete structure as fundamental. A finite candidate grid of possible magnets is supplied, each with a prescribed maximum dipole magnitude and orientation constraint, and magnets are activated iteratively according to the reduction they provide in a surface-error metric, typically the $\mathbf B\cdot\mathbf n$ residual on the plasma boundary~\cite{Kaptanoglu2023}. The greedy structure ensures rapid convergence and allows for explicit engineering constraints, such as binary placement, minimum separations, or orientation restrictions.




Two enhancements, arbitrary-vector (ArbVec) selection and backtracking, further improve robustness~\cite{Kaptanoglu2023,Hammond2022QUASAR,Hammond2024CPC}. In the original ``basis'' GPMO implementation, each candidate site is scored using a \emph{discrete}, user-prescribed set of allowed moment directions (a small fixed menu, commonly taken to be a Cartesian basis such as $\pm\hat{\mathbf x},\pm\hat{\mathbf y},\pm\hat{\mathbf z}$, or whichever orientations a particular holder design permits). ArbVec modifies only this orientation step: instead of choosing one direction from that discrete menu, the algorithm treats the user-specified directions as a spanning set, forms the corresponding precomputed surface-response fields, and then chooses the best \emph{linear combination} (subject to the maximum dipole strength) so the trial dipole moment can point in a continuously optimized direction within that user-defined span at each site. Backtracking then revisits earlier greedy commitments and can remove (or swap) previously chosen magnets if later steps reveal that a different active subset yields a lower surface error. In the present work, the rigid-remanence inner loop used by GPMOmr is the ArbVec variant throughout (and ArbVec is likewise used in all GPMOmr backtracking runs), with macromagnetic refinement applied only after the ArbVec winner is committed.

For later use we denote by $\Gamma_t$ the \emph{active set} of grid locations after $t$ greedy steps, i.e.\ the subset of candidate sites that host magnets at iteration $t$.

\section{Analysis: Deriving a macromagnetic model for device scale}
Having recalled the micromagnetic description, we now coarse-grain to a block-level macromagnetic model appropriate for the millimetre-scale bricks used in MUSE and similar arrays, and derive the equilibrium condition used throughout the rest of the paper.

\subsection{The problem setup}
We focus on permanent-magnet arrays at room temperature, where the thermodynamic state is effectively fixed and temperature-dependent effects such as spin fluctuations or Curie transitions can be neglected. Each block is modeled as a hard magnet: the crystallographic easy axis $\hat{\mathbf u}$ is set during fabrication and remains locked to the lattice under all relevant operating fields. The remanent state is $\mathbf M_{\mathrm{rem}}=M_{\mathrm{rem}}\hat{\mathbf u}$, with small perturbations arising from finite susceptibility. For MUSE specifically, the PM arrays and holders operate at ambient laboratory conditions, and the macromagnetic parameters used here correspond to room-temperature values consistent with the device report~\cite{Qian2023}.


For rare-earth compounds such as Nd--Fe--B, the longitudinal and transverse susceptibilities $\chi_\parallel,\chi_\perp$ are both small but nonzero. In practice $\chi_\perp$ may reach values of order $10^{-1}$, producing observable canting when strong demagnetizing fields are present, while $\chi_\parallel$ is typically an order of magnitude smaller~\cite[Chs.~6--8]{Jiles2016}. Thus, although the rigid-remanent approximation has been invoked in PM stellarator synthesis, a more realistic macromagnetic model fixes the easy axis $\hat{\mathbf u}$ but allows $\mathbf M$ to deviate slightly from $M_{\mathrm{rem}}\hat{\mathbf u}$ via both tilt (rotation away from $\hat{\mathbf u}$) and magnitude changes.

\subsection{Finite-$\mu$: isotropic versus anisotropic response}
The simplest constitutive closure is the isotropic finite-$\mu$ model, in which the magnetization responds to the local internal field as
\begin{equation}
\mathbf M = \chi\,\mathbf H + \mathbf M_{\mathrm{rem}},
\label{eq:isotropic_M}
\end{equation}
with $\chi$ a scalar susceptibility. In this case the permeability is $\mu=\mu_0(1+\chi)$ and the susceptibility tensor is simply
\begin{equation}
\boldsymbol\chi_{\mathrm{iso}}
= \chi \mathbf I
= \begin{bmatrix}
\chi & 0 & 0 \\
0 & \chi & 0 \\
0 & 0 & \chi
\end{bmatrix}.
\label{eq:chi_iso}
\end{equation}
This closure suffices when the material is magnetically isotropic, e.g.\ polycrystalline samples without texture.

For crystalline hard magnets, however, the response is anisotropic with respect to the easy axis $\hat{\mathbf u}$. The susceptibility becomes a rank-2 tensor,
\begin{equation}
\boldsymbol\chi
= \chi_\parallel\,\hat{\mathbf u}\hat{\mathbf u}^{\!\top}
 + \chi_\perp\!\left(\mathbf I-\hat{\mathbf u}\hat{\mathbf u}^{\!\top}\right),
\label{eq:chi_aniso}
\end{equation}
so that fields parallel and perpendicular to $\hat{\mathbf u}$ excite distinct responses. Explicitly, in the local basis$\{\hat{\mathbf e}_1,\hat{\mathbf e}_2,\hat{\mathbf u}\}$ one obtains
\begin{equation}
\boldsymbol\chi_{\mathrm{aniso}}
= \begin{bmatrix}
\chi_\perp & 0 & 0 \\
0 & \chi_\perp & 0 \\
0 & 0 & \chi_\parallel
\end{bmatrix}.
\label{eq:chi_matrix}
\end{equation}
Equation~\eqref{eq:chi_matrix} makes clear that deviations along $\hat{\mathbf u}$ merely stretch or compress the remanent magnitude, while transverse fields induce small tilts of $\mathbf M$. This formalism also admits generalizations with distinct transverse responses ($\chi_{\perp1},\chi_{\perp2}$), but for the present device-scale study we restrict attention to the uniaxial case and use constant room-temperature Nd--Fe--B values throughout, taking $\mu_{\parallel}=1.05$ along $\hat{\mathbf u}$ and $\mu_{\perp}=1.15$ perpendicular to $\hat{\mathbf u}$ (equivalently $\chi_{\parallel}=0.05$ and $\chi_{\perp}=0.15$). These values reflect the near-ideal, weakly permeable response of sintered Nd--Fe--B and are consistent with the MUSE material model and reported measurements of anisotropic permeability~\cite{Katter2005,Qian2023,Jiles2016,Bjoerk2021}.

These tensor closures provide the foundation for constructing macromagnetic energy functionals: the isotropic case reduces to a quadratic penalty in \mbox{$\|\mathbf M-\mathbf M_{\mathrm{rem}}\|^2$}, while the anisotropic case resolves longitudinal and transverse contributions separately, as in Eq.~\eqref{eq:chi_aniso}.

\subsection{Micromagnetics vs.\ macromagnetics: exchange length and fixed easy axis}
In micromagnetics, the magnetization field $\mathbf m(\mathbf r)=\mathbf M/M_s$ is treated as a continuous unit vector whose gradients carry an energetic penalty through the exchange term $g_{\mathrm{ex}}=A_{\mathrm{ex}}\|\nabla \mathbf m\|^2$ [Eq.~\eqref{eq:g_ex}]. The role of exchange is to suppress spatial inhomogeneities of $\mathbf m$ within each grain, ensuring that neighboring spins remain aligned over nanometric distances. This balance is quantified by the \emph{exchange length}, obtained by comparing exchange energy density $A_{\mathrm{ex}}/\ell^2$ with magnetostatic self-energy $\mu_0 M_s^2$~\cite{DonahueMiltat2007,ExlSuessSchrefl2013}:
\begin{equation}
\ell_{\mathrm{ex}}
=\sqrt{\frac{2\,A_{\mathrm{ex}}}{\mu_0 M_s^2}} .
\label{eq:lex}
\end{equation}
For Nd$_2$Fe$_{14}$B magnets at room temperature, typical values are $M_s \simeq 1.3\times 10^6\ \mathrm{A/m}$ ($\mu_0 M_s \approx 1.61$\,T) and $A_{\mathrm{ex}}\approx 8$--$12.5$\,pJ/m~\cite{ExlSuessSchrefl2013}. Substituting these into Eq.~\eqref{eq:lex} gives $\ell_{\mathrm{ex}} \approx 3\,\mathrm{nm}$, many orders of magnitude smaller than the millimetre-scale bricks and gaps relevant to device arrays. In practice, intra-grain exchange enforces uniform magnetization within each block, and $g_{\mathrm{ex}}$ can be neglected at device scale.

By contrast, the spacing between discrete blocks is determined by engineering tolerances. In the MUSE design, for example, a minimum edge gap $d_m=1$\,mm is enforced, with additional clearance $d_s=2d_m$ on both sides, yielding a physical gap of at least $4$\,mm between holders~\cite{Qian2023}. This is six orders of magnitude larger than $\ell_{\mathrm{ex}}$, showing that inter-block coupling is purely magnetostatic. The macromagnetic regime is therefore defined by: (i) uniform bricks, (ii) negligible exchange, and (iii) fixed crystallographic axes.

A similar simplification applies to anisotropy. The uniaxial term $g_{\mathrm{ani}}=-K_u(\mathbf m\!\cdot\!\hat{\mathbf u})^2$ [Eq.~\eqref{eq:g_ani}] energetically locks $\mathbf m$ to the easy axis $\hat{\mathbf u}$. In sintered Nd--Fe--B, grains are aligned during fabrication in multi-Tesla fields, and the resulting anisotropy field $H_\text{ani}=2K_u/(\mu_0 M_s)$ remains several Tesla thereafter~\cite[Chs.~6--8]{Jiles2016,Herbst1991R2Fe14B}. Since this far exceeds the operating fields in stellarators, $\hat{\mathbf u}$ is effectively fixed. Thus, macromagnetics discards $g_{\mathrm{ex}}$, treats $\hat{\mathbf u}$ as immobile, and introduces finite-$\mu$ effects only through the tensor susceptibility $\boldsymbol\chi$ [Eq.~\eqref{eq:chi_aniso}], which acts at the block scale rather than through spatial gradients.


\subsection{Work required to change $\mathbf M$ and the limits of Brown's equation}
Classical micromagnetics imposes Brown's equilibrium condition
\begin{equation}
\mathbf m \times \mathbf H_{\mathrm{eff}}=0,
\qquad \|\mathbf m\|=1,
\label{eq:Brown_discussion}
\end{equation}
derived from Landau--Lifshitz dynamics [Eq.~\eqref{eq:LLG}]. This enforces rigid length, $|\mathbf m|=1$. However, once finite susceptibilities are admitted, $\mathbf M$ can vary both longitudinally and transversely, breaking the unit-sphere constraint. To describe such variable-length magnetizations at the block scale we work with a Helmholtz free-energy functional $F[\mathbf M]$ rather than the micromagnetic Gibbs functional $G[\mathbf m]$ above. The reversible magnetic work governing changes in $\mathbf M$ is
\begin{equation}
dF=\mu_0\,\mathbf H\cdot d\mathbf M ,
\label{eq:dF_discussion}
\end{equation}
as derived rigorously in Appendix~\ref{app:reversible-work} [Eq.~\eqref{eq:A15}]. Combining Eq.~\eqref{eq:dF_discussion} with the constitutive law $\mathbf M-\mathbf M_{\mathrm{rem}}=\boldsymbol\chi\,\mathbf H$ and integrating along a straight path in magnetization space leads to the quadratic penalty
\begin{equation}
f_\chi(\mathbf M)=\frac{\mu_0}{2}\left[\frac{\|\mathbf M_\perp\|^2}{\chi_\perp}
+\frac{(M_\parallel-M_{\mathrm{rem}})^2}{\chi_\parallel}\right],
\label{eq:fchi_discussion}
\end{equation}
shown explicitly in Appendix~\ref{app:quadratic-form} [Eq.~\eqref{eq:B5}]. This term augments the classical exchange, demagnetizing, anisotropy, and Zeeman energies and yields the full macromagnetic free-energy functional $F[\mathbf M]$ given in Eq.~\eqref{eq:Fstar}. Within this framework the rigid-length constraint is removed and stationarity is imposed by $\delta F/\delta\mathbf M = 0$ rather than Brown's constraint on $\mathbf m$.

\subsection{Equilibrium condition and linear system for finite--$\mu$ blocks}



With the augmented functional $F[\mathbf M]$ including the susceptibility penalty
[Eq.~\eqref{eq:fchi_discussion}], equilibrium corresponds to the vanishing
of the effective field
\begin{equation}
\mathbf H_{\mathrm{eff}}
= \mathbf H_a + \mathbf H_d
 - \boldsymbol\chi^{-1}\bigl(\mathbf M-\mathbf M_{\mathrm{rem}}\bigr),
\label{eq:Heff_analysis}
\end{equation}
where $\mathbf H_a$ is the applied field and
$\mathbf H_d=-\sum_j \underline{\underline{N}}_{ij}\mathbf M_j$ is the
demagnetizing field from all blocks through the sample-shape tensor
$\underline{\underline{N}}_{ij}$~\cite{Bjoerk2021,Jiles2016}.

\emph{Geometric dependence:} $\underline{\underline N}_{ij}$ depends only on the shapes, orientations, and relative locations of bodies $i$ and $j$; it is independent of their magnetizations. In a candidate grid, sites that are not selected correspond to the absence of material (no body), so they carry no magnetization and have no rows/columns in the active system. In practice we therefore assemble only the $\underline{\underline N}_{ij}$ blocks for the current active set $\Gamma_t$, although a precomputation over the full candidate geometry is also possible.

The stationarity condition $\mathbf H_{\mathrm{eff}}=0$ therefore gives
\begin{equation}
\boldsymbol\chi^{-1}\!\left(\mathbf M_i-\mathbf M_{\mathrm{rem}}\hat{\mathbf u}_i\right)
= \mathbf H_a - \sum_{j=1}^N \underline{\underline N}_{ij}\mathbf M_j ,
\label{eq:stationary_analysis}
\end{equation}
for each block $i=1,\dots,N$. This is the unconstrained (variable-length) analogue of Brown's equation: with the quadratic susceptibility term in $F$ the rigid $|\mathbf m|=1$ constraint is relaxed, and stationarity reduces to $\mathbf H_{\mathrm{eff}}=0$ (see Appendices~\ref{app:reversible-work}--\ref{app:quadratic-form}). Multiplying Eq.~\eqref{eq:stationary_analysis} by $\boldsymbol\chi_i$ and rearranging terms produces
\begin{equation}
\mathbf M_i + \sum_{j=1}^N \boldsymbol\chi_i\,\underline{\underline N}_{ij}\mathbf M_j
= M_{\mathrm{rem}}\hat{\mathbf u}_i + \boldsymbol\chi_i\,\mathbf H_a .
\label{eq:linear_block}
\end{equation}

Assembling the Cartesian components of all $N$ magnets into a single
vector $\mathbf M\in\mathbb R^{3N}$, Eq.~\eqref{eq:linear_block} becomes
a global linear system
\begin{equation}
\underbrace{\bigl[\delta_{ij}\mathbf I_3+\boldsymbol\chi_i\,\underline{\underline N}_{ij}\bigr]_{i,j=1}^N}_{\displaystyle A}
\,\mathbf M
= \underbrace{\bigl[M_{\mathrm{rem}}\hat{\mathbf u}_i+\boldsymbol\chi_i\mathbf H_a\bigr]_{i=1}^N}_{\displaystyle b} ,
\label{eq:Am_b}
\end{equation}
which has the standard algebraic form
\begin{equation}
A\,\mathbf M = b .
\end{equation}

Equation~\eqref{eq:Am_b} is the macromagnetic equilibrium condition:
given remanence, susceptibilities, easy-axis orientations, and the
demagnetizing tensors, the equilibrium magnetizations follow from a
single linear solve. This reformulation is exact within the
uniform-block approximation, and provides a computationally efficient
pathway to incorporate finite-$\mu$ effects into device-scale PM
design and optimization.

\section{Computational Macromagnetics}
We now discuss how the macromagnetic equilibrium system is solved numerically and how its structure motivates the use of Krylov methods on large PM arrays.

The macromagnetic equilibrium condition derived in Eq.~\eqref{eq:Am_b} has the algebraic form
\begin{equation}
A\,\mathbf M = b ,
\end{equation}
where $\mathbf M\in\mathbb{R}^{3N}$ is the concatenated magnetization vector of all $N$ degrees of freedom on the PM grid (one three-component magnetization per grid location that can host a block), $b$ encodes the remanent contributions and applied field, and the system matrix $A$ is defined blockwise as
\begin{equation}
A_{ij} = \delta_{ij}\mathbf I_3 + \boldsymbol\chi_i\,\underline{\underline N}_{ij}.
\end{equation}
Here $\underline{\underline N}_{ij}$ is the demagnetization tensor between grid locations $i$ and $j$, and $\boldsymbol\chi_i$ the anisotropic susceptibility tensor of location $i$ if it hosts material. By construction $\underline{\underline N}_{ij}$ is symmetric, and if all blocks share the same scalar susceptibility ($\boldsymbol\chi_i=\chi\mathbf I$), then $A$ is also symmetric. In the more general case of block-dependent or anisotropic $\boldsymbol\chi_i$, the weighting differs across rows and columns, and $A$ is no longer symmetric. This rules out classical conjugate-gradient methods and motivates Krylov subspace solvers such as GMRES~\cite{SaadSchultz1986,Saad2003}.


\subsection{Cost and sparsity structure}
The dominant operation in GMRES is the application of $A$ to a trial vector $\mathbf v\in\mathbb R^{3N}$. Writing $\mathbf v=(\mathbf v_1,\dots,\mathbf v_N)$ with $\mathbf v_i\in\mathbb R^3$, one has
\begin{equation}
(A\mathbf v)_i = \mathbf v_i + \sum_{j=1}^N \boldsymbol\chi_i\,\underline{\underline N}_{ij}\mathbf v_j .
\end{equation}
If all candidate sites are active and all pairwise demagnetization blocks $\underline{\underline N}_{ij}$ are retained, this is an $O(N^2)$ operation per matrix--vector application. In practice, however, only $m=|\Gamma_t|$ of the $N$ candidate locations host material at a given refinement step, and we assemble $A$ only over this active subset, giving a $3m\times 3m$ system. For MUSE-scale grids with $m\sim 10^4$, the resulting dense $O(m^2)$ matrix--vector multiply is still tractable on modern workstations.

On very large arrays, further savings are possible by truncating the demagnetization tensor beyond a prescribed distance or drop tolerance, so that only interactions within a geometrically defined neighborhood of each magnet are retained~\cite{Bjoerk2021}. In that case each row of $A$ couples only to $O(z)$ other magnets, where $z$ is the typical number of neighbors within the truncation radius, and matrix--vector products can be implemented using a neighbor-list representation with effective cost $O(m z)$ per iteration. In the present work we use the dense active-set formulation since $m$ remains moderate.

\subsection{Discussion of method choice}
The combination of a dense but geometrically structured demagnetization operator, a nonsymmetric system matrix, and moderate accuracy requirements makes Krylov subspace solvers the natural choice for macromagnetic equilibrium. They allow re-use of previous iterates as initial guesses (important in optimization loops), can be preconditioned using block-diagonal or low-rank approximations to the demagnetization tensor, and scale tractably to the $m\sim 10^4$--$10^5$ active blocks relevant for device-scale arrays, especially when the underlying demagnetization operator is stored in an active-set or neighbor-sparse form.

\section{Macromagnetic postprocessing}
\label{sec:macromag-post}

We now apply the macromagnetic model developed above to the published MUSE PM grid~\cite{Qian2023}. The candidate grid used in the MUSE design contains $N_{\rm cand}=9736$ possible magnet locations arranged in ``towers'' between the circular TF coils and the vacuum vessel~\cite{Qian2023}. Here a ``tower'' denotes the stack of cuboid candidate tiles generated by the FAMUS ``towering'' procedure~\cite{Zhu2020,Hammond2020GeomPM}, in which points on the plasma boundary are projected outward along the local surface normal to form lines of potential magnet locations. The synthesis and assessment of permanent-magnet layouts through the surface normal field metric $\mathbf B\cdot\mathbf n$ follows the standard pipeline used by FAMUS and related tools~\cite{Zhu2020,Hammond2020GeomPM}, while our device-scale magnetic response is modeled with uniform block susceptibilities consistent with hard-magnet phenomenology~\cite{Jiles2016,Herbst1991R2Fe14B,Sagawa1984NdFeB}. Unless otherwise stated, configuration-space averages below are taken over a single unique half-period wedge of the stellarator so that discrete symmetries do not trivially drive the mean to zero.

\subsection{Grid symmetry and discretization effects}
Most stellarators are designed with two discrete symmetries: field-period symmetry and stellarator symmetry. However, in practice any finite-size rectangular magnets that intersect a symmetry plane will slightly break these assumptions. 
Breaking of the discrete symmetries also comes from the towering procedure used to construct the MUSE candidate grid: towers are generated by extruding along surface normals, and the resulting voxelization does not enforce that opposing towers are cut into identical volumes when they intersect the symmetry planes. In particular, the overlap of PM volumes with the symmetry planes leads to asymmetric clipping of the voxels, exactly analogous to the overlap effects discussed for coils in topology-optimization formulations of inverse magnetostatics~\cite{Kaptanoglu2024InverseMagnetostatics}. In the continuous limit the plasma surface and coil set are stellarator symmetric, but the finite-volume PM grid only approximates these symmetries. As a result, any calculation that does not explicitly enforce the symmetry operations will exhibit small but finite symmetry-breaking contributions at the level of the discretization error.

\subsection{Relation to micromagnetic analysis}

The MUSE paper performed a closely related study of finite permeability using the tile-based micromagnetic framework MagTense~\cite{Bjoerk2021,Qian2023}. In that assessment, MagTense solved the magnetostatic micromagnetic problem at the tile scale, capturing both effective tilts and magnitude changes of the local magnetization due to anisotropic permeability and demagnetizing fields. However, the self-consistent part of the calculation included only magnet–magnet interactions; fields from the TF coils were not coupled into the micromagnetic solve. Our macromagnetic formulation can, by contrast, incorporate both magnet–magnet and coil–magnet couplings on an equal footing via $\mathbf H_a$, enabling direct comparison between coil-inclusive and coil-free macromagnetic corrections for MUSE.

For quantitative comparisons we use the same surface-normal-field least-squares objective $f_B$ as in the classical GPMO formulation of~\cite{Kaptanoglu2023}. On a discrete surface grid $\{\mathbf x_q,\mathbf n_q,w_q\}$ with target normal field $\mathbf B_{\rm targ}\cdot\mathbf n_q$, we write
\begin{equation}
f_B = \sum_q w_q\,\bigl[(\mathbf B\cdot\mathbf n)_q - (\mathbf B_{\rm targ}\cdot\mathbf n)_q\bigr]^2 ,
\label{eq:fB_def}
\end{equation}
so that $f_B$ is the weighted sum of squared normal-field residuals on the target surface. This definition is equivalent, up to an overall factor of $1/2$ and the explicit quadrature weights $w_q$, to the matrix form $f_B = \tfrac{1}{2}\|A m - b\|_2^2$ used in~\cite{Kaptanoglu2023}, where $A$ and $b$ encode the normal field on the plasma boundary. All $f_B$ values reported in this section are evaluated on a surface with $n_\phi=n_\theta=1024$.

\subsection{Tilt and magnitude changes}

Solving the macromagnetic equilibrium system [Eq.~\eqref{eq:Am_b}] with these parameters yields blockwise magnetizations $\mathbf M_i$ that deviate slightly from their remanent targets $M_{\mathrm{rem}}\hat{\mathbf u}_i$, in line with expectations for small but nonzero anisotropic susceptibilities in hard magnets~\cite{Jiles2016}. Two channels of deviation are distinguished: an effective tilt angle
\begin{equation}
\Delta\theta_i=\arccos\!\bigl(\hat{\mathbf u}_i\cdot\mathbf M_i/|\mathbf M_i|\bigr),
\end{equation}
and a change in magnitude
\begin{equation}
\Delta|M_i|=|\mathbf M_i|-M_{\mathrm{rem}}.
\end{equation}

We consider three postprocessing cases on the fixed MUSE GPMO layout: (i) the uncoupled rigid-remanence solution, (ii) a macromagnetic solution with magnet--magnet coupling only, and (iii) a fully coupled macromagnetic solution with both magnet--magnet and coil coupling. For the magnet--magnet-only case we obtain
\begin{align}
\langle \Delta\theta \rangle_{\rm mm} &= 1.06^{\circ}, \\
\max \Delta\theta_{\rm mm} &= 3.16^{\circ}, \\
\langle |\Delta M| \rangle_{\rm mm} &= 1.59\times 10^{4}\ \mathrm{A/m}, \\
\max |\Delta M|_{\rm mm} &= 4.02\times 10^{4}\ \mathrm{A/m},
\end{align}
while for the fully coupled (magnet--magnet + coil) case we find
\begin{align}
\langle \Delta\theta \rangle_{\rm mc} &= 1.16^{\circ}, \\
\max \Delta\theta_{\rm mc} &= 3.81^{\circ}, \\
\langle |\Delta M| \rangle_{\rm mc} &= 1.59\times 10^{4}\ \mathrm{A/m}, \\
\max |\Delta M|_{\rm mc} &= 4.03\times 10^{4}\ \mathrm{A/m}.
\end{align}
Using $M_{\mathrm{rem}}\simeq 1.2\times 10^6\ \mathrm{A/m}$ for Nd--Fe--B~\cite{Herbst1991R2Fe14B,Jiles2016}, these correspond to
\begin{align}
\frac{\langle |\Delta M| \rangle_{\rm mm}}{M_{\mathrm{rem}}} &\approx 1.3\%\,, &
\frac{\max |\Delta M|_{\rm mm}}{M_{\mathrm{rem}}} &\approx 3.3\%\,, \\
\frac{\langle |\Delta M| \rangle_{\rm mc}}{M_{\mathrm{rem}}} &\approx 1.3\%\,, &
\frac{\max |\Delta M|_{\rm mc}}{M_{\mathrm{rem}}} &\approx 3.4\% .
\end{align}
The finite-$\mu$ response thus produces both tilts and magnitude changes. Including coil coupling shifts the statistics only slightly (by $\lesssim 0.1^{\circ}$ in tilt and at the $10^{-3}$ level in the relative magnitude changes) but systematically increases the typical deviation from remanence: coils push the magnets a bit further away from the rigid-remanence state while leaving the overall pattern qualitatively unchanged.

\subsection{Impact on $\mathbf B\cdot\mathbf n$ residuals}

The plasma-facing figure of merit is the surface normal field $\mathbf B\cdot\mathbf n$, the same metric minimized during stage-two synthesis in prior work~\cite{Zhu2020,Qian2023}. We therefore evaluate the difference
\begin{equation}
\Delta(\mathbf B\cdot\mathbf n)
= (\mathbf B\cdot\mathbf n)_{\mathrm{unc}}
  - (\mathbf B\cdot\mathbf n)_{\mathrm{mc}},
\label{Bn_delta}
\end{equation}
between the fully coupled macromagnetic field (magnet--magnet + coil, subscript ``mc'') and the uncoupled rigid-remanence solution (subscript ``unc'').

\begin{figure}[t]
    \centering
    \includegraphics[width=\columnwidth]{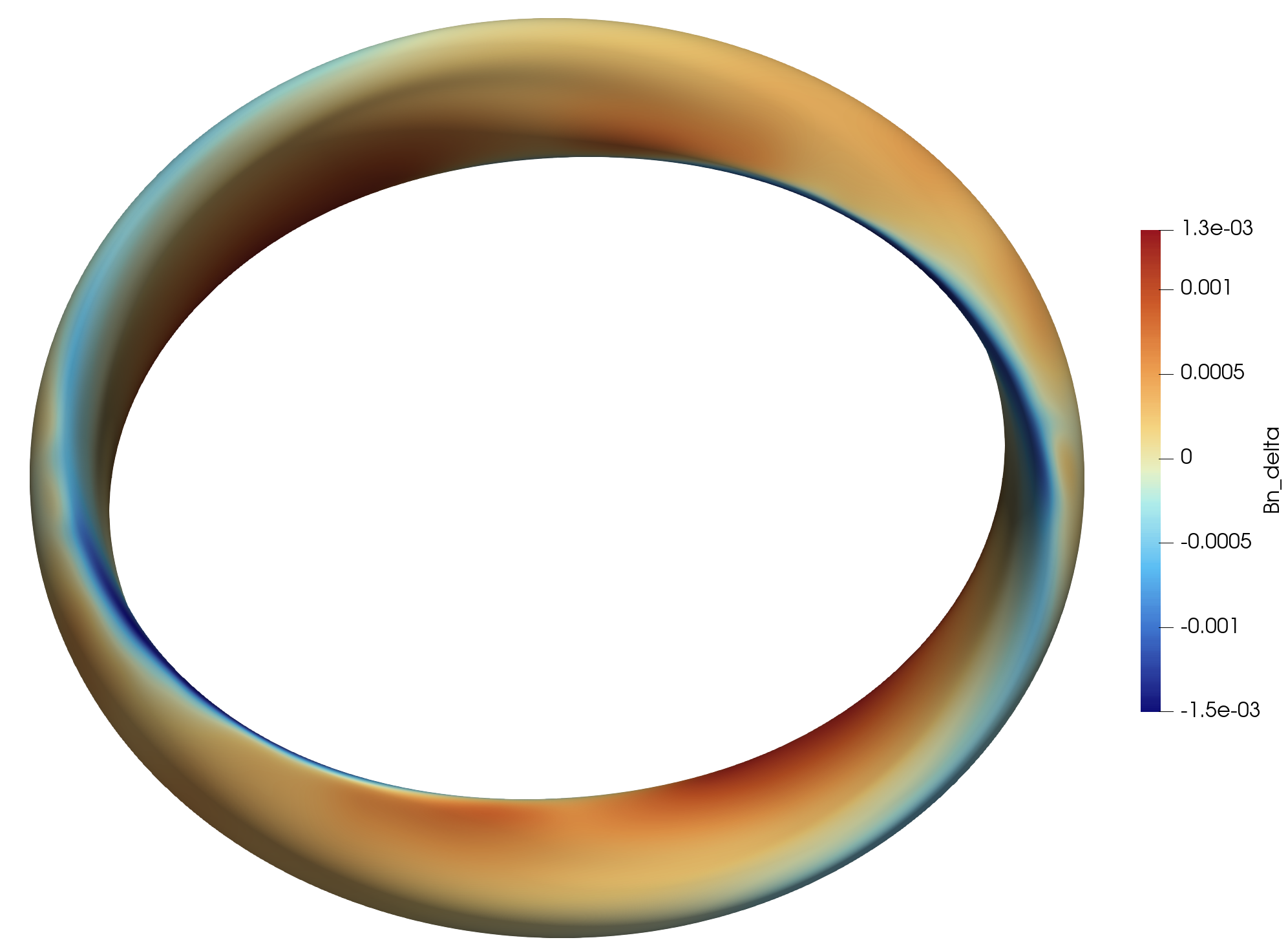}
    \caption{Difference in surface normal field $\Delta(\mathbf B\cdot\mathbf n)$ between the uncoupled rigid-remanence solution and the fully coupled macromagnetic solution on the MUSE plasma boundary. The color bar spans approximately $[-1.5,1.3]\times10^{-3}$~T, corresponding to a peak fractional change of $1.00\%$ relative to the $0.15$~T target field. Small asymmetries in the extrema reflect discretization-induced violations of stellarator symmetry in the PM grid rather than any large macromagnetic effect.}
    \label{fig:muse-bn-delta}
\end{figure}

Figure~\ref{fig:muse-bn-delta} shows this difference field on the full MUSE plasma surface. The color scale ranges from approximately $-1.5\times10^{-3}$~T to $+1.3\times10^{-3}$~T, and the corresponding peak fractional change is
\[
\max\frac{|\Delta(\mathbf B\cdot\mathbf n)|}{0.15\ \mathrm{T}} = 1.00\%,
\]
using the stage-two MUSE target of $0.15$~T on the plasma boundary as a reference~\cite{Qian2023}. Over a unique half-period wedge we find
\begin{equation}
\frac{\langle \Delta(\mathbf B\cdot\mathbf n)\rangle}{0.15\ \mathrm{T}} = 4.35\times 10^{-4}.
\end{equation}
In a perfectly symmetric discretization, with symmetry operations enforced exactly, the extrema of the color bar would be equal and opposite and the mean would vanish by construction. The slight imbalance visible in Fig.~\ref{fig:muse-bn-delta} (upper and lower bounds differing by about $15\%$) is entirely consistent with the discretization asymmetries in the PM grid discussed above, together with the fact that the macromagnetic solve is carried out on the full geometry rather than on a single fundamental domain.

\begin{figure*}[t]
    \centering
    \includegraphics[width=\textwidth]{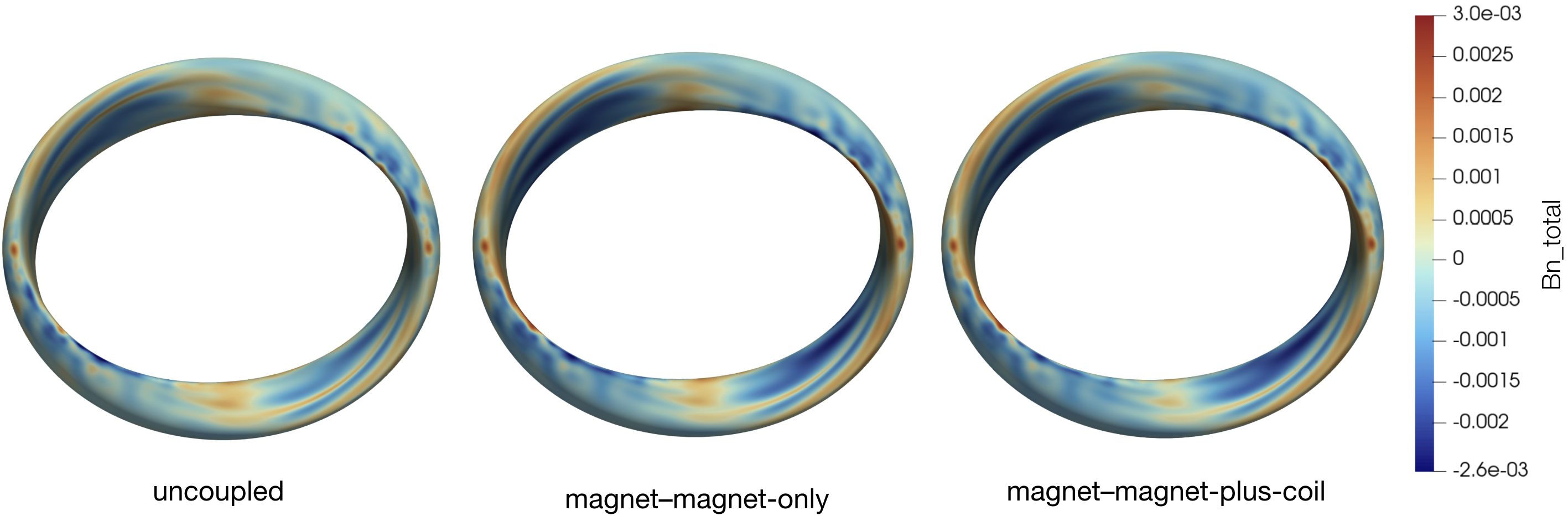}
    \caption{Surface $\mathbf B\cdot\mathbf n$ residual on the MUSE plasma boundary for three postprocessed fields: (left) uncoupled rigid-remanence solution, (center) macromagnetic solution with magnet--magnet coupling only, and (right) macromagnetic solution with both magnet--magnet and coil coupling. The patterns are qualitatively similar in all cases; macromagnetic and coil coupling primarily modulate the amplitude of existing structures rather than creating new defects. The color scale spans $[-2.6,3.0]\times10^{-3}$~T.}
    \label{fig:muse-bn-total}
\end{figure*}

To place these differences in the context of the overall error budget, Fig.~\ref{fig:muse-bn-total} shows the absolute surface $\mathbf B\cdot\mathbf n$ residual for all three postprocessed fields. From left to right, the panels display the uncoupled rigid-remanence solution, the macromagnetic solution with magnet--magnet coupling only, and the macromagnetic solution with both magnet--magnet and coil coupling. All three panels use the same color scale, which is dominated by existing error structures at the order $10^{-3}$~T level. The macromagnetic corrections do not create new large-scale defects; they slightly redistribute and smear the existing residuals. Comparing the center and right panels shows that including coil coupling modestly amplifies some of the residual bands, consistent with the slightly larger magnitude changes reported above.

Over the unique half-period wedge, we denote the uncoupled, magnet--magnet-only, and magnet--magnet-plus-coil cases by the subscripts ``unc'', ``mm'', and ``mc'', respectively. The average $\mathbf B\cdot\mathbf n$ residuals are
\begin{align}
\langle (\mathbf B\cdot\mathbf n)_{\rm unc}\rangle &= 2.94\times 10^{-5}\ \mathrm{T}, \\
\langle (\mathbf B\cdot\mathbf n)_{\rm mm}\rangle &= -1.44\times 10^{-5}\ \mathrm{T}, \\
\langle (\mathbf B\cdot\mathbf n)_{\rm mc}\rangle &= -3.59\times 10^{-5}\ \mathrm{T}.
\end{align}
All three averages are of order $10^{-5}$~T, confirming that the mean residual remains small and that macromagnetics primarily reshapes the distribution of errors rather than introducing a large bias.

The effect on the integrated squared-flux objective $f_B$, however, is significantly larger. Evaluating Eq.~\eqref{eq:fB_def} on the high-resolution $n_\phi=n_\theta=1024$ surface grid yields
\begin{align}
f_B^{\rm unc} &= 8.40\times 10^{-8}, \\
f_B^{\rm mm}  &= 1.86\times 10^{-7}, \\
f_B^{\rm mc}  &= 1.90\times 10^{-7}.
\end{align}
In other words, while $\Delta(\mathbf B\cdot\mathbf n)$ remains at the $\sim1\%$ level pointwise, the global squared-flux objective for this fixed PM layout increases by more than a factor of two when finite-permeability effects are included, with a small additional increase when coil coupling is also considered. This tension, namely that the $\mathbf B\cdot\mathbf n$ maps remain visually similar while $f_B$ changes by $O(1)$, is central for interpreting the design-stage results in Sec.~\ref{sec:results}.

\subsection{Interpretation}

These postprocessing results confirm that macromagnetic corrections for MUSE are small in a local, pointwise sense but non-negligible when measured by the global design metric. The finite-$\mu$ response produces degree-level tilts and few-percent changes in magnetization magnitude, with coils slightly enhancing the typical deviation from remanence. On the plasma surface, these corrections translate into smooth, percent-level changes in the normal field, without introducing new large-scale defects and with only modest symmetry breaking attributable to the discretized PM grid and towering-induced volume asymmetries. At the same time, the squared-flux objective $f_B$ for the \emph{same} PM layout increases by more than a factor of two once macromagnetics is taken into account.

In this sense, the postprocessing study reveals a tension that is not obvious from the surface plots alone. The $\mathbf B\cdot\mathbf n$ maps before and after macromagnetic refinement remain visually similar and differ at only the $\sim 1\%$ level, yet the integrated error measure $f_B$ is substantially larger. 
Consequently, a layout that appears well optimized under the rigid-remanence model can be noticeably suboptimal once finite permeability and demagnetizing interactions are included.

Taken together, these findings suggest a useful separation of roles. For a given towered PM layout---such as the one obtained from the FAMUS-based optimization and tower construction in MUSE---the idealized, rigid-remanent evaluation provides a good first approximation to a low-$f_B$ configuration, while macromagnetic postprocessing serves as a quantitative diagnostic of how vulnerable that layout is to finite-$\mu$ effects. The fact that modest blockwise tilts and magnitude shifts can change $f_B$ by order unity then motivates going beyond pure postprocessing and developing an iteration scheme that \emph{builds} macromagnetic corrections into the synthesis step itself, so that one searches directly for PM grids that achieve comparably low $f_B$ even after finite permeability is taken into account.

\section{Macromag GPMO}
We next describe how the macromagnetic equilibrium solve is coupled to the greedy optimization loop, leading to the GPMOmr algorithm.

\subsection{Integrating the macromagnetic solve into GPMO}

A straightforward but prohibitively expensive coupling would evaluate the macromagnetic equilibrium [Eq.~\eqref{eq:Am_b}] for \emph{every} candidate at each greedy step, i.e.\ embed a full finite-$\mu$ solve inside the scoring loop. Let $N_{\rm cand}$ denote the total number of candidate sites and $m$ the number of active magnets at a given step. If $m$ magnets are already active and $N_{\rm cand}-m$ candidates remain, this implies $O(N_{\rm cand}-m)$ solves on systems of size $3(m{+}1)\times 3(m{+}1)$ per iteration. Even with Krylov methods (Sec.~III), the per-solve cost scales as $O(k\,m^2)$ (each $A$--vector product is $O(m^2)$; $k$ iterations), so a single iteration would cost $O\bigl((N_{\rm cand}-m)\,k\,m^2\bigr)$. Summing $m=1\ldots N_{\rm cand}$ yields an overall cost proportional to $kN_{\rm cand}^4$ (more precisely, to $k\sum_{m=1}^{N_{\rm cand}} (N_{\rm cand}-m)m^2 \sim kN_{\rm cand}^4/12$)~\cite{Kaptanoglu2023,SaadSchultz1986,Saad2003}. This is impractical for device-scale arrays, as modern stellarator designs such as PM4STELL can involve on the order of $3.5\times10^4$ individual magnet placements~\cite{Zhu2022PM4Stell}.

Instead we use a ``winner-only'' refinement strategy that mirrors the structure of the original GPMO algorithm~\cite{Kaptanoglu2023,Zhu2020,Hammond2022QUASAR,Hammond2024CPC}. Candidates are first scored using the rigid-remanence ArbVec criterion, exactly as in the classical algorithm, so that the inner loop remains a fast, purely dipole-based calculation with no macromagnetic solve. The best-scoring candidate is then committed and the active set $\Gamma_t$ is updated. Only at this point is the macromagnetic model invoked: a \emph{single} finite-$\mu$ equilibrium solve is performed on the current active set, using the restriction of Eq.~\eqref{eq:Am_b} to indices in $\Gamma_t$. The resulting equilibrium magnetizations $\{\mathbf M_i\}_{i\in\Gamma_t}$ are used to update the cached $\mathbf B\cdot\mathbf n$ residual, so that all subsequently placed magnets see a background field that is already consistent with the macromagnetic response of the committed blocks.


On large problems it is convenient to call the macromag solve only every $k_{\rm mm}$ greedy iterations (for example $k_{\rm mm}=50$). In that case, intermediate steps reuse the last available macromag solution and initialize newly added magnets with their remanent moment. In the MUSE runs reported here we used $k_{\rm mm}=50$; tests with smaller $k_{\rm mm}$ values showed qualitatively similar $f_B$ histories at higher computational cost, so we do not show them separately. In practice this ``subsampled'' refinement retains the qualitative behavior of the fully coupled scheme while keeping the cost close to that of the classical GPMO implementation.

\subsection{Incremental matrix assembly and Krylov solve}
Let $m_\Gamma =|\Gamma_t|$ be the number of active blocks at refinement step $t$. The macromagnetic system at that step is the restriction of Eq.~\eqref{eq:Am_b} to the indices in $\Gamma_t$, i.e.\ a $3m\times 3m$ submatrix of the full operator acting on the vector of block magnetizations $\mathbf M_{\Gamma_t}\in\mathbb{R}^{3m}$. When a new block $p$ is added to the active set, we do not rebuild this system from scratch. Instead, we append a single $3\times 3$ row and column associated with $p$, and extend the right-hand side by one additional three-component entry. Only the demagnetization tensors $\underline{\underline N}_{ip}$ and $\underline{\underline N}_{pp}$ required for this update are evaluated, so the assembly cost per refinement step grows smoothly with $m$.


In practice we carry the current matrix $A_m$, right-hand side $b_m$, and solution $\mathbf M_{\Gamma_t}$ forward between macromag refinement steps, reusing the previously assembled blocks whenever new magnets are added. Because each refinement modifies $A_m$ only incrementally, the linear solve difficulty typically changes only gradually as $m$ increases. Here ``smoothly with $m$'' means that, for a fixed Krylov tolerance, the required iteration count does not exhibit abrupt jumps when one appends a new $3\times 3$ block; instead it tends to drift slowly (often remaining nearly constant over wide ranges of $m$ in our examples), so the dominant growth in refinement cost comes from the increasing cost of the matrix--vector products for a $3m\times 3m$ dense block system rather than from rapidly worsening conditioning. This incremental update strategy keeps the algebraic structure simple and allows the macromag refinement to be slotted into the greedy loop with minimal additional overhead.

\paragraph*{Field handling.}
The applied field $\mathbf H_a$ is evaluated once at all candidate centers (for example by Biot--Savart from the coils) and cached. Each refinement step then simply gathers $\{\mathbf H_a(\mathbf r_i)\}_{i\in\Gamma_t}$ for the active indices, avoiding repeated coil evaluations inside the greedy loop.

\paragraph*{Demagnetization evaluation.}
The prism demagnetization tensors $\underline{\underline N}_{ij}$ are evaluated only for those index pairs $(i,j)$ that appear in the current active system $A_m$ and in the new row/column associated with a freshly added block. No global table of $N_{ij}$ over the full candidate grid is ever assembled. This dynamic strategy keeps the assembly cost proportional to $m$ per refinement, while the Krylov cost grows smoothly with $m$. Importantly the memory storage costs are $\mathcal{O}(m^2)$ instead of $\mathcal{O}(N_{\rm cand}^2)$, which would be prohibitive for these problems without the use of high-memory computing devices.

\begin{algorithm}[t]
    \caption{GPMOmr: greedy permanent-magnet optimization with macromagnetic refinement}
    \label{alg:gpmo-macromag}
    \begin{algorithmic}[1]
    \Require target surface $\{ \mathbf x_q, \mathbf n_q, w_q \}$, coil field $\mathbf B_{\mathrm{coil}}$, 
    candidate grid $\{ \mathbf r_i, m_i^{\max}, \hat{\mathbf u}_i \}$,
    susceptibility tensors $\{ \boldsymbol\chi_i \}$, maximum greedy iterations $K$
    \Ensure set of active magnets $\Gamma_K$ and equilibrium magnetizations $\mathbf M_{\Gamma_K}$
    
    \Statex Initialize $\Gamma_0 \gets \emptyset$, $A_0 \gets []$, $b_0 \gets []$
    \For{$t \gets 1$ \textbf{to} $K$}
        \Statex \textbf{Greedy scoring with rigid remanence}
        \For{each candidate $p \notin \Gamma_{t-1}$}
          \State Temporarily activate $p$ (rigid remanence, ArbVec) and compute the trial change in $\mathbf B\cdot\mathbf n$ on the surface
          \State Evaluate updated surface error $\mathcal J(\Gamma_{t-1}\cup\{p\})$
        \EndFor
        \State Select winner $p^\star \gets \arg\min_p \mathcal J(\Gamma_{t-1}\cup\{p\})$
        \State Update active set $\Gamma_t \gets \Gamma_{t-1}\cup\{p^\star\}$
        \Statex \textbf{Macromagnetic refinement of the committed set (every $k_{\rm mm}$ steps)}
        \If{$t \bmod k_{\rm mm} = 0$}
          \State Form $(A_t,b_t)$ for $\Gamma_t$ by appending a $3\times 3$ row and column for $p^\star$ to $(A_{t-1},b_{t-1})$ according to Eq.~\eqref{eq:Am_b}
          \State Solve $A_t \mathbf M_{\Gamma_t} = b_t$ with a Krylov method (e.g.\ GMRES) to obtain updated equilibrium magnetizations
          \State Update cached $\mathbf B\cdot\mathbf n$ residual using $\mathbf M_{\Gamma_t}$
        \Else
          \State Initialize $\mathbf M_{p^\star}$ at remanence and reuse the previous macromag solution on $\Gamma_{t-1}$
        \EndIf
    \EndFor
    \State \Return $\Gamma_K$, $\mathbf M_{\Gamma_K}$
    \end{algorithmic}
\end{algorithm}

Algorithm~\ref{alg:gpmo-macromag} summarizes the overall GPMOmr scheme used in the numerical experiments below.

\section{Results}
\label{sec:results}
We now compare classical (uncoupled) GPMO with GPMOmr on the MUSE grid, first without and then with backtracking, focusing on the squared-flux objective and the resulting magnetization patterns.

\subsection{Setup and parameter choices}
All comparisons in this section use the published MUSE permanent-magnet grid~\cite{Qian2023}. We run both the classical (uncoupled) GPMO and GPMOmr on the same candidate set without backtracking, using a surface resolution of $n_\phi = n_\theta = 64$ for the $\mathbf B\cdot\mathbf n$ error evaluation and a maximum of $K_{\max} = 20{,}000$ greedy iterations. The target surface and coil configuration are identical to those in the MUSE design, so that any differences between the runs arise solely from the inclusion of macromagnetic corrections inside the greedy loop. For GPMOmr we scan three choices of macromagnetic refinement interval, $k_{\rm mm}\in\{1,25,50\}$, so that the finite-$\mu$ equilibrium is recomputed after every $k_{\rm mm}$ greedy placements.

\subsection{Squared-flux error history}
\begin{figure}[t]
    \centering
    \includegraphics[width=0.48\textwidth]{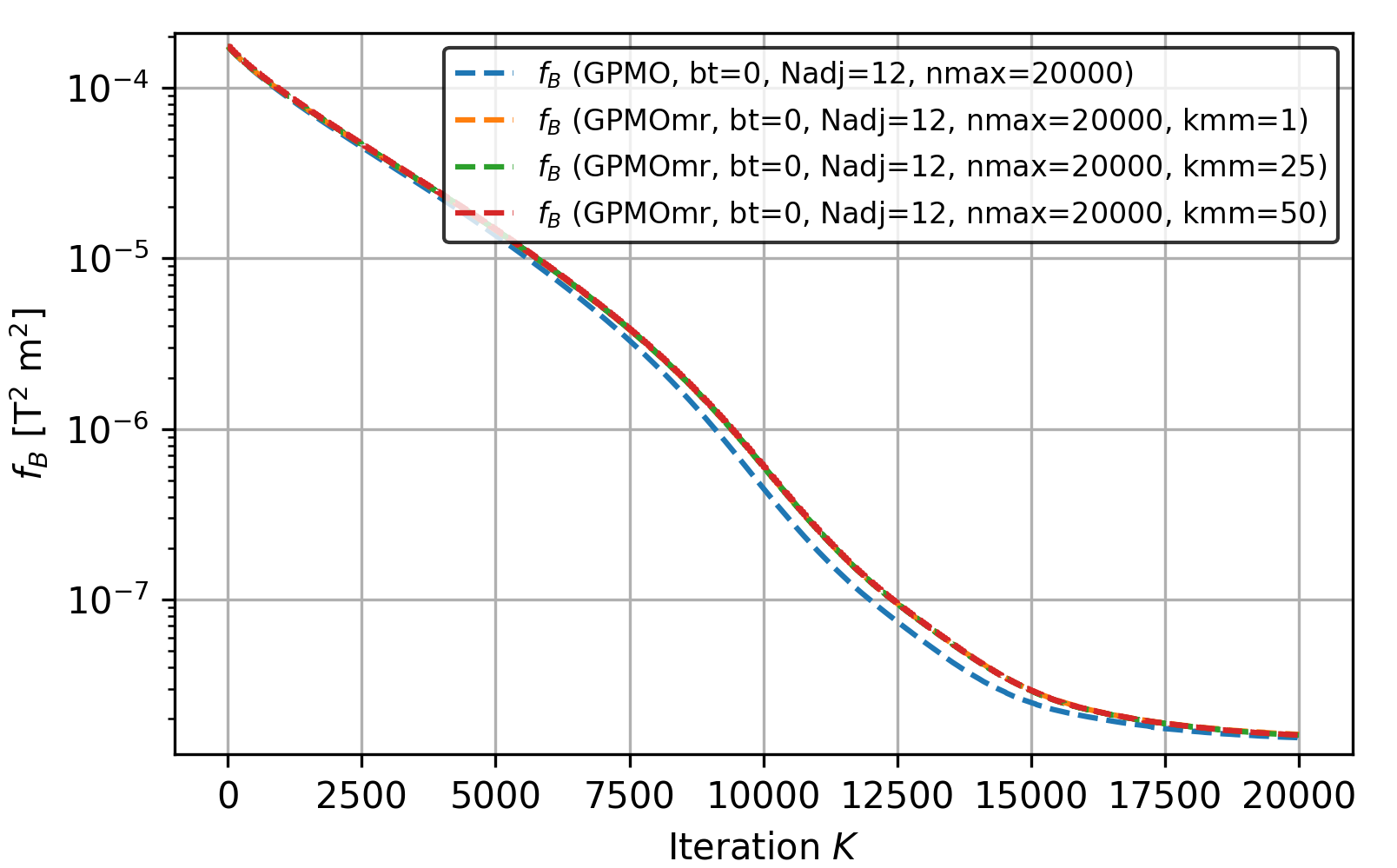}
    \caption{Squared-flux error $f_B$ versus greedy iteration $K$ for classical GPMO (uncoupled) and GPMOmr on the MUSE PM grid (no backtracking). The three GPMOmr curves correspond to macromagnetic refinement intervals $k_{\rm mm}=1,25,50$ and lie almost on top of one another, indicating that the choice of $k_{\rm mm}$ has little effect on the $f_B(K)$ history while strongly affecting the number of macromagnetic solves.}
    \label{fig:mse-history}
\end{figure}

Figure~\ref{fig:mse-history} shows the evolution of the squared-flux objective $f_B$ as a function of greedy iteration $K$ for classical GPMO and for the three GPMOmr runs with $k_{\rm mm}=1,25,50$. All four histories lie very close to one another on the logarithmic scale and exhibit smooth, monotone convergence over the entire run. The three GPMOmr curves for different $k_{\rm mm}$ choices are visually almost indistinguishable: at fixed $K$ they track the same path within the line thickness and differ from the classical GPMO curve only at the level of a few percent in $f_B$.

These histories show that including macromagnetic feedback inside the scoring loop modifies the detailed path of the optimization but does not change the achievable error level for this grid and material model. They also indicate that the choice of refinement interval mainly affects computational cost rather than convergence quality: taking $k_{\rm mm}=1$ recomputes the macromagnetic equilibrium after every greedy placement and is therefore much more expensive in wall-clock time, while $k_{\rm mm}=25$ and $k_{\rm mm}=50$ achieve essentially the same $f_B(K)$ history with far fewer macromagnetic solves. In the remainder of this section we therefore use $k_{\rm mm}=50$ as the default refinement interval for GPMOmr, as it provides a good balance between refinement of the magnetization field and overall runtime.

\subsection{Magnetization patterns on the MUSE grid}
\begin{figure*}[t]
    \centering
    \includegraphics[width=\textwidth]{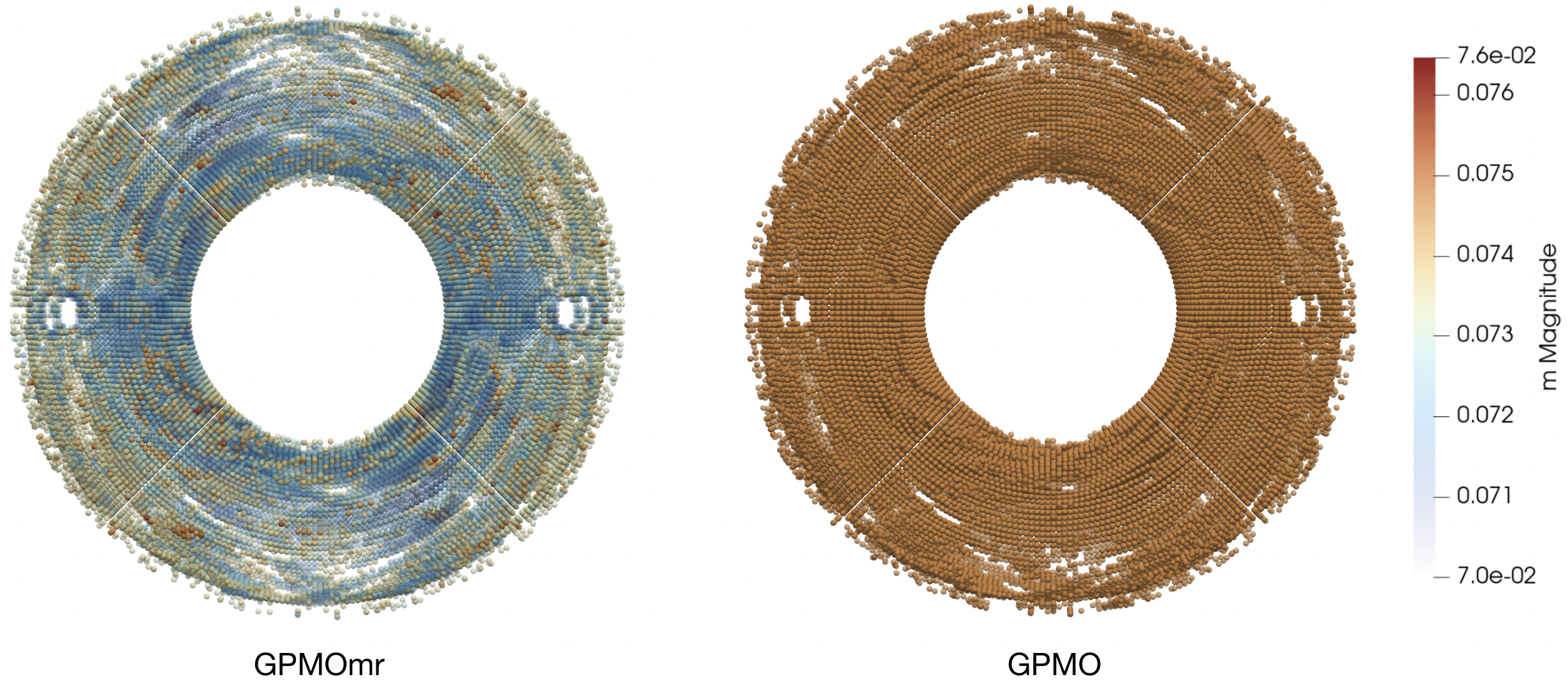}
    \caption{Glyph plots of the final magnetization patterns on the MUSE PM grid without backtracking. Left: GPMOmr, including finite-$\mu$ corrections through the macromagnetic solve. Right: classical uncoupled GPMO. The uncoupled solution exhibits nearly uniform magnitudes, while the macromag solution shows substantial spatial variation driven by local demagnetizing fields.}
    \label{fig:pm-grid}
\end{figure*}

To see how the two algorithms use the available candidate sites, we plot the final magnetization vectors on the MUSE grid for both runs. Figure~\ref{fig:pm-grid} shows side-by-side permanent magnet grid plots of the classical GPMO solution and the GPMOmr solution.

In the classical uncoupled case, activated magnets are essentially rigid: each block is either off or carries a dipole  aligned with its allowed axis and near a single remanent magnitude. The resulting pattern is visually uniform, with all active sites displaying identical magnetization length.

The GPMOmr configuration is visibly different. Because each refinement step enforces macromagnetic equilibrium for the active set, local demagnetizing fields produce small reductions or enhancements in the magnetization magnitude and slight tilts away from the nominal axis. As a result, clusters of magnets in strongly interacting regions show noticeably reduced moments, while more isolated blocks remain closer to the remanent value. The two arrays therefore achieve almost the same surface error using magnetization fields that differ both in support and in local amplitude and direction.

\subsection{Pointwise magnetization differences}
\begin{figure*}[t]
    \centering
    \includegraphics[width=\linewidth]{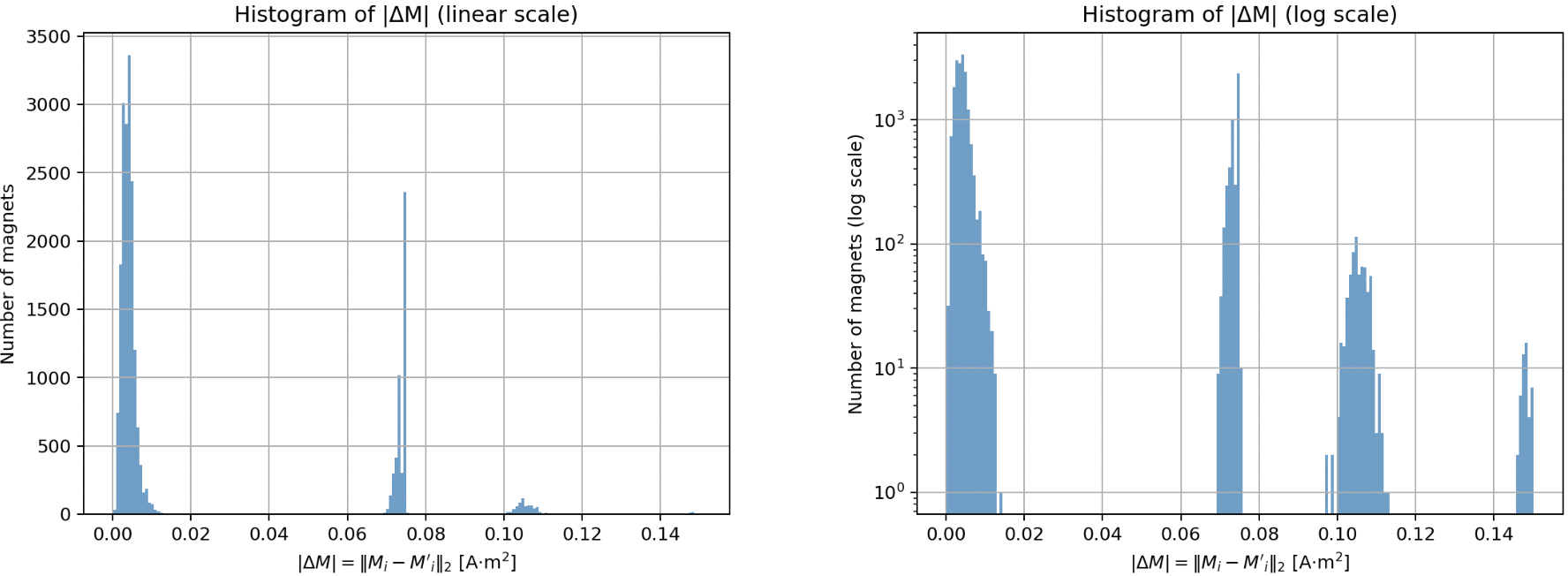}
    \caption{Histograms of the pointwise dipole-moment difference $\Delta M_i$ between the macromagnetic-refinement and classical runs on the MUSE grid (no backtracking). Four clear peaks appear: near zero (almost identical blocks), near the single-dipole cap $m_{\max}$ (blocks active in only one run), near $\sqrt{2}\,m_{\max}$ (approximately orthogonal placements), and near $2m_{\max}$ (rare anti-parallel cases).}
    \label{fig:deltaM}
\end{figure*}

To quantify the differences between the two magnetization fields, we compute the pointwise Euclidean norm
\begin{equation}
\begin{aligned}
\Delta M_i &= \bigl\|\mathbf m_i^{\mathrm{GPMOmr}} - \mathbf m_i^{\mathrm{GPMO}}\bigr\|_2, \\
m_{\max} &\equiv V\,M_{\mathrm{rem}}.
\end{aligned}
\end{equation}
for all positions $i$ on the grid, where $\mathbf m_i$ denotes the final \emph{dipole-moment} vector at grid site $i$ (in A$\cdot$m$^{2}$). Here $V$ is the prism volume and $M_{\mathrm{rem}}$ is the remanent magnetization; since the GPMO runs use identical prisms at all grid sites (constant cube dimensions), $m_{\max}=V M_{\mathrm{rem}}$ provides the characteristic single-dipole scale for the corresponding run and is used throughout the paper when interpreting $\Delta M$ histograms. Figure~\ref{fig:deltaM} shows histograms of $\Delta M_i$ on both a linear and a logarithmic $y$-scale.


The distribution is sharply structured and separates into four distinct buckets. First, there is a large peak very close to $\Delta M\approx 0$, corresponding to sites where the two runs agree almost exactly. These blocks either remain inactive in both runs or, if active, experience only small angular or amplitude changes under macromagnetic refinement. Second, a cluster appears near $\Delta M \approx m_{\max}$, corresponding to positions that are magnetized in one run but not the other: for example, a site that carries a nearly saturated dipole in the GPMOmr solution but is left empty by the classical run (or vice versa). Third, there is a clear peak near $\Delta M \approx \sqrt{2}\,m_{\max}$, consistent with blocks whose dipole moments differ by roughly $90^\circ$ between the runs, so that the vector difference has length $\sqrt{2}\,m_{\max}$ when both runs place a dipole of comparable magnitude but along nearly orthogonal directions. Finally, a smaller peak occurs near $\Delta M \approx 2m_{\max}$, corresponding to rare cases where the two runs place nearly anti-parallel dipoles at the same grid site, giving $|\Delta M|\approx 2m_{\max}$.

In absolute terms, $70.5\%$ of all candidate sites are inactive in both runs. Among the remaining sites that are active in at least one of the two runs, $76.3\%$ fall in the near-zero bucket, $20.5\%$ lie near $m_{\max}$, $2.9\%$ lie near $\sqrt{2}\,m_{\max}$, and $0.2\%$ lie near $2m_{\max}$.


This structure shows that, while most sites are either unchanged or experience only mild corrections, a nontrivial subset of blocks is used in fundamentally different ways by the two algorithms. Nevertheless, as the next subsection shows, the resulting differences in the plasma-facing $\mathbf B\cdot\mathbf n$ field remain very small.

\subsection{Surface \texorpdfstring{$\mathbf B\cdot\mathbf n$}{B·n} comparison}

We now examine the surface normal field $\mathbf B\cdot\mathbf n$, which is the quantity minimized in the GPMO objective. At the level of global statistics, the two patterns are almost indistinguishable: both achieve comparable RMS error and similar peak deviations, consistent with the small difference in final $f_B$ reported above. This close agreement is expected, since both classical GPMO and GPMOmr minimize the same squared-flux objective on the same target surface with the same coil background; once $f_B$ has been driven to a similarly low value, the residual $\mathbf B\cdot\mathbf n$ distribution is strongly constrained and only modest redistributions of error are possible.

The spatial structure of the residuals is illustrated in Appendix~\ref{app:surface-bn-no-bt}, which collects the full-surface $\mathbf B\cdot\mathbf n$ plots for both algorithms without backtracking. The two distributions match pointwise to within a few milliTesla over most of the surface, and the locations of dominant error structures are preserved. Visually, the macromag solution introduces only subtle changes: in some regions, elongated streaks of residual $\mathbf B\cdot\mathbf n$ present in the classical solution coalesce into more compact islands, and certain sharp features appear slightly smoothed. This behavior is consistent with the internal redistribution of magnetization magnitudes and tilts seen in Fig.~\ref{fig:pm-grid}: the macromagnetic solve re-balances local fields so that gradients in $\mathbf B\cdot\mathbf n$ are softened, while the overall plasma-facing normal field remains essentially the same as in the classical rigid-remanence design.

\subsection{GPMOmr with backtracking}
\label{sec:results-backtracking}


In a second series of runs we enabled backtracking in both the classical and GPMOmr algorithms, using the same MUSE PM grid, material model, and error evaluation as above but with a maximum of $K_{\max}=25{,}000$ greedy iterations and a backtracking depth of $n_{\mathrm{bt}}=200$. During each backtracking stage the algorithm examines a local neighborhood of size $N_{\mathrm{neighbors}}=12$ around each candidate and removes magnets whose dipole moments are nearly antiparallel to close neighbors---configurations that tend to produce locally canceling fields and slow convergence. The angular cutoff for this test was $\theta_{\mathrm{thresh}}=\pi - 5^\circ$. We use this slightly relaxed threshold, rather than a strict $\theta_{\mathrm{thresh}}=\pi$, to ensure that genuinely canceling, nearly antiparallel configurations are still detected even when the macromagnetic equilibrium solves introduce small deviations from perfect $180^\circ$ alignment. All other parameters (surface resolution, coil configuration, candidate set, and stopping criteria) were held fixed.

\begin{figure}[t]
    \centering
    \includegraphics[width=0.48\textwidth]{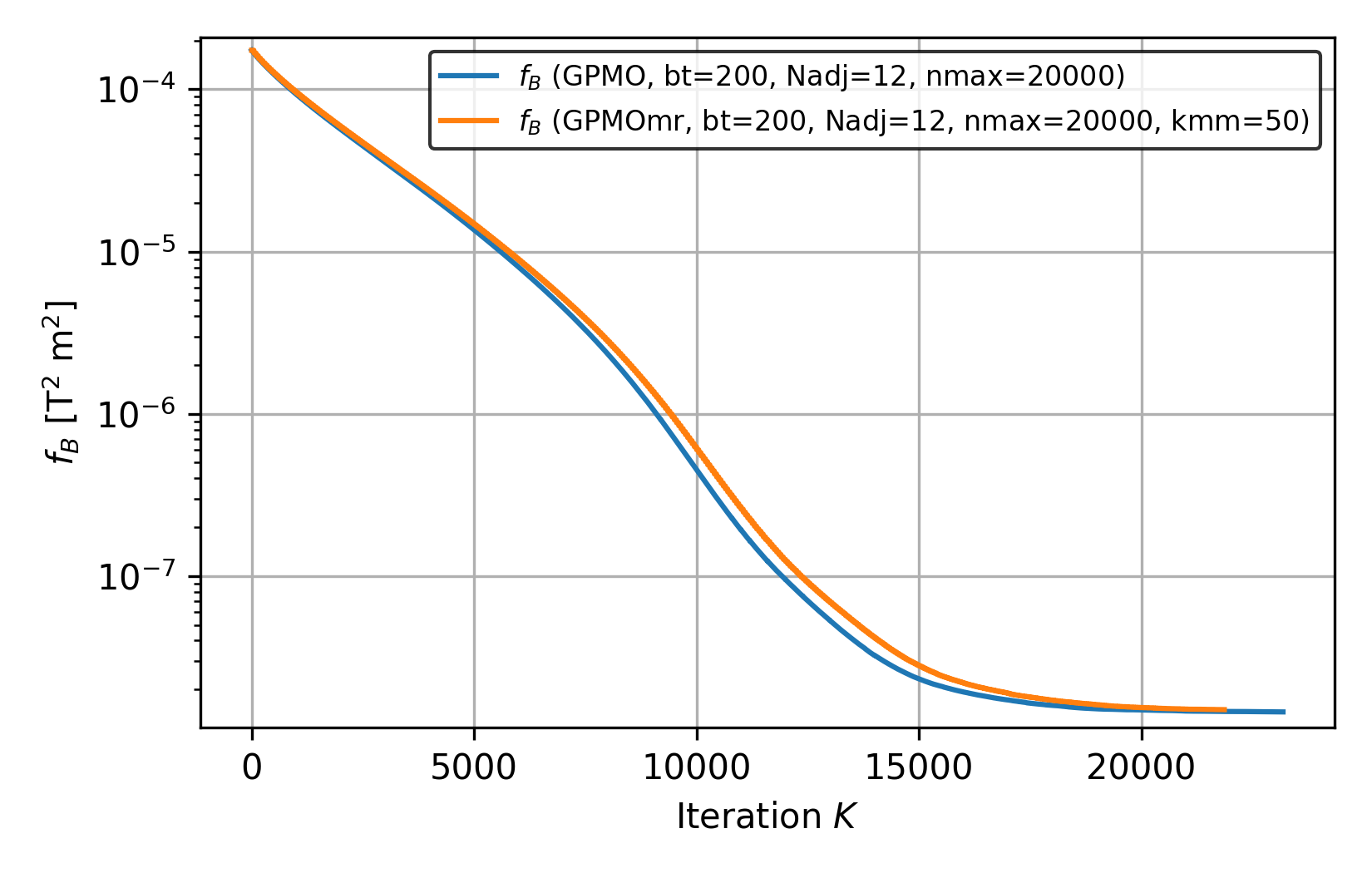}
    \caption{Squared-flux error $f_B$ versus greedy iteration $K$ for classical GPMO (uncoupled) and GPMOmr on the MUSE PM grid with backtracking enabled ($n_{\mathrm{bt}}=200$). Both runs converge more rapidly than in Fig.~\ref{fig:mse-history}. The classical run attains a slightly lower final error and remains marginally ahead for most of the history, but the difference between the two algorithms is still only at the few-percent level.}
    \label{fig:mse-history-bt}
\end{figure}

Figure~\ref{fig:mse-history-bt} shows the resulting squared-flux histories. Compared with the no-backtracking case in Fig.~\ref{fig:mse-history}, both curves drop more rapidly in the first $10^4$ iterations, as expected when poorly performing early placements can be removed or replaced. The classical GPMO run remains slightly ahead of the macromag run throughout most of the history and converges marginally faster near the end of the optimization. For $K\gtrsim 1.5\times 10^4$ the GPMOmr curve begins to flatten, indicating that the macromag run is effectively hitting its magnet-count ceiling earlier in iteration space; subsequent backtracking stages primarily reshuffle or locally adjust existing blocks rather than adding new ones, so further reductions in $f_B$ are modest. The uncoupled run continues to make small gains out to $K_{\max}$, which accounts for its slightly lower final error.

High-resolution evaluations of the final states give
\begin{equation}
\begin{aligned}
f_B^{\mathrm{GPMO}}     &\approx 1.45\times 10^{-8}, \\
f_B^{\mathrm{GPMOmr}}   &\approx 1.50\times 10^{-8},
\end{aligned}
\end{equation}
so backtracking lowers the absolute error for both algorithms but the gap between them remains only at the few-percent level. Thus, including macromagnetic feedback inside a backtracking GPMO loop does not significantly change the minimum surface-error level that can be reached on this grid.

The detailed structure of the magnetization differences and plasma-facing fields in the backtracking runs closely mirrors the no-backtracking case. Histograms of the pointwise difference $\Delta M_i$ exhibit the same four peaks discussed in Fig.~\ref{fig:deltaM}, and the corresponding surface $\mathbf B\cdot\mathbf n$ maps show the same error structures at nearly identical locations and amplitudes, with GPMOmr again only softening some localized peaks.

\subsection{Extending permanent magnet stellarators to higher field strengths}
\label{sec:results-gb50uh}
One motivation for this work is to extend permanent magnet stellarator design to higher field strengths and stronger magnet-magnet and magnet-coil coupling.
To probe how sensitive device-scale PM optimization is to realistic NdFeB grades under stronger applied fields, we repeat a representative high-resolution GPMOmr run on the published MUSE grid. The coil currents are scaled linearly so that the average field on the major radius is $B_0 \approx 0.5$~T and permanent magnets experience peak fields on the order of $\approx 1$~T. To avoid demagnetization concerns, we consider grain-boundary-diffused NdFeB grade GB50UH magnets using room-temperature datasheet values~\cite{Arnold2021GBDNeoCatalog}. Importantly, at $20^\circ$C these magnets demagnetize at $\approx 3.08$~T. If the magnets are cooled to e.g. liquid nitrogen temperatures, the demagnetization strengths are much larger still ($\approx 3.95$~T), making them potentially usable in substantially higher-field stellarator designs (although very large volumes of this class of NdFeB magnets would be required to generate the high fields). For the N52 NdFeB magnets used by MUSE, the magnets fully demagnetize at $\approx 1.1$~T, preventing use in stellarators with higher field strengths. We set the remanent induction to $B_{\max}=1.410$~T (typical for the GB50UH magnets at $20^\circ$C). All other modeling choices are kept fixed: the same candidate geometry and tower discretization are used, the same anisotropic permeability model is retained with $\mu_{\mathrm{ea}}=1.05$ and $\mu_{\mathrm{oa}}=1.15$, and coil coupling enters through the applied-field term $\mathbf H_a$ in the macromagnetic equilibrium solve. Both coil coupling and demagnetizing interactions are enabled. As in our other material comparisons, the MUSE candidate grid stores reference dipole-moment scales corresponding to the baseline material. For GB50UH we therefore rescale the per-site maximum dipole moment in proportion to $B_{\max}$ while keeping the cell volume fixed, i.e.
\begin{equation}
m_{\max}^{(\mathrm{GB50UH})}
=
m_{\max}^{(\mathrm{base})}\,
\frac{B_{\max}^{(\mathrm{GB50UH})}}{B_{\max}^{(\mathrm{base})}},
\end{equation}
which enforces the correct dipole-moment scale for the new material without changing the underlying cube dimensions.


The GB50UH run uses the same device-scale optimization settings as in the large MUSE demonstrations, with $n_\phi=64$ and $K=5\times 10^4$ greedy iterations. Backtracking is performed every 200 iterations with $N_{\mathrm{adjacent}}=12$ and a limit of $4\times 10^4$ nonzero magnets. Macromagnetic refinement is performed every $k_{\mathrm{mm}}=100$ iterations; this cadence balances refinement coarseness against computational cost at the larger magnet counts used in this run.

\begin{figure*}[t]
    \centering
    \includegraphics[width=\textwidth]{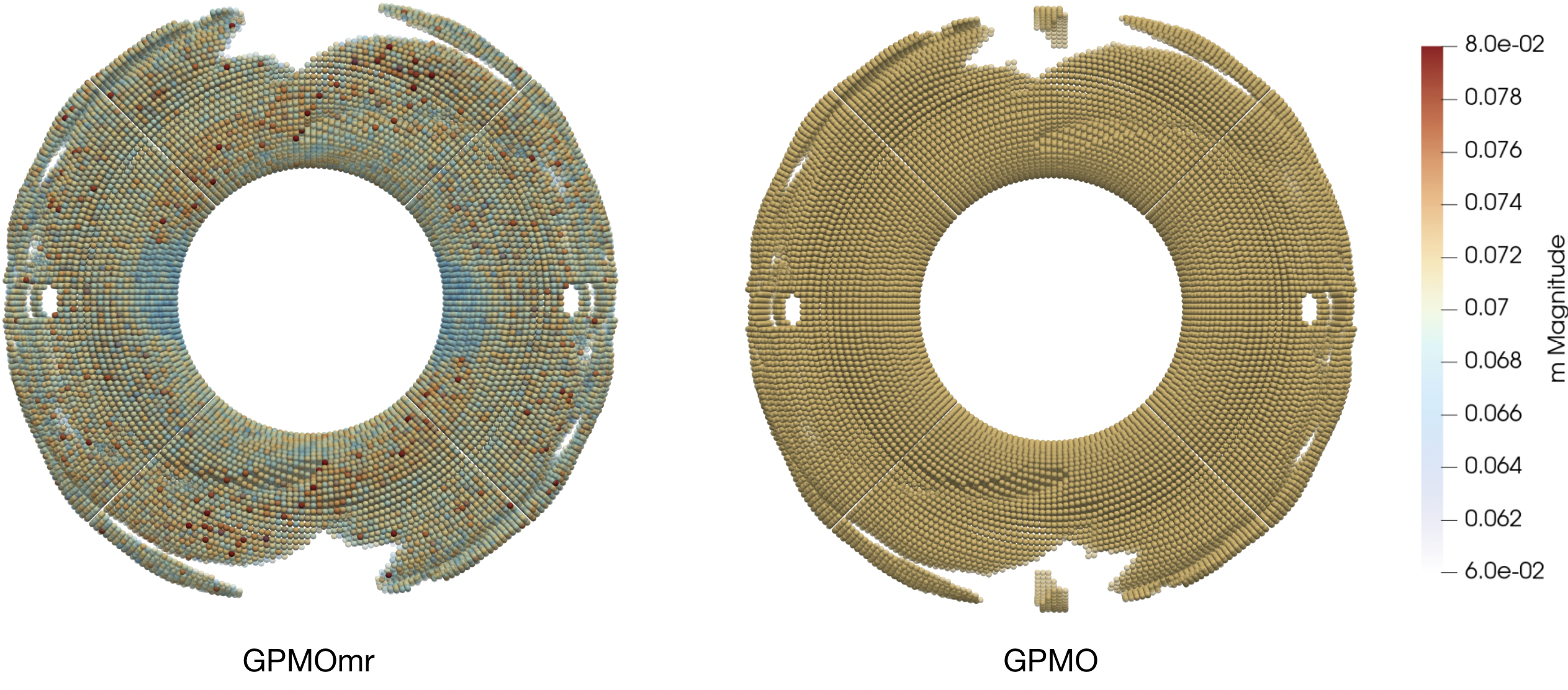}
    \caption{Glyph plots of the final magnetization patterns on the MUSE PM grid for the GB50UH substitution study. The remanent induction is set to $B_{\max}=1.410$~T at room temperature~\cite{Arnold2021GBDNeoCatalog}. Colors indicate the dipole-moment magnitude $m$, with the plotted range spanning $[6,8]\times 10^{-2}$~A$\cdot$m$^{2}$. The maximum plotted magnitude is $\approx 8\times 10^{-2}$~A$\cdot$m$^{2}$, and no grid-wide demagnetization is observed in this device-scale configuration.}
    \label{fig:gb50uh-pm-grid}
\end{figure*}

Figure~\ref{fig:gb50uh-pm-grid} shows the resulting magnetization patterns for the classical and macromagnetic runs. The overall magnitude scale is reduced relative to the baseline material, consistent with the reduced remanent induction and the fixed prism volume used in these runs. No device-scale collapse of magnetization magnitudes is observed. The plotted dipole-moment magnitudes span $[6,8]\times 10^{-2}$~A$\cdot$m$^{2}$, with a maximum of $\approx 8\times 10^{-2}$~A$\cdot$m$^{2}$.

\begin{figure}[t]
    \centering
    \includegraphics[width=\columnwidth]{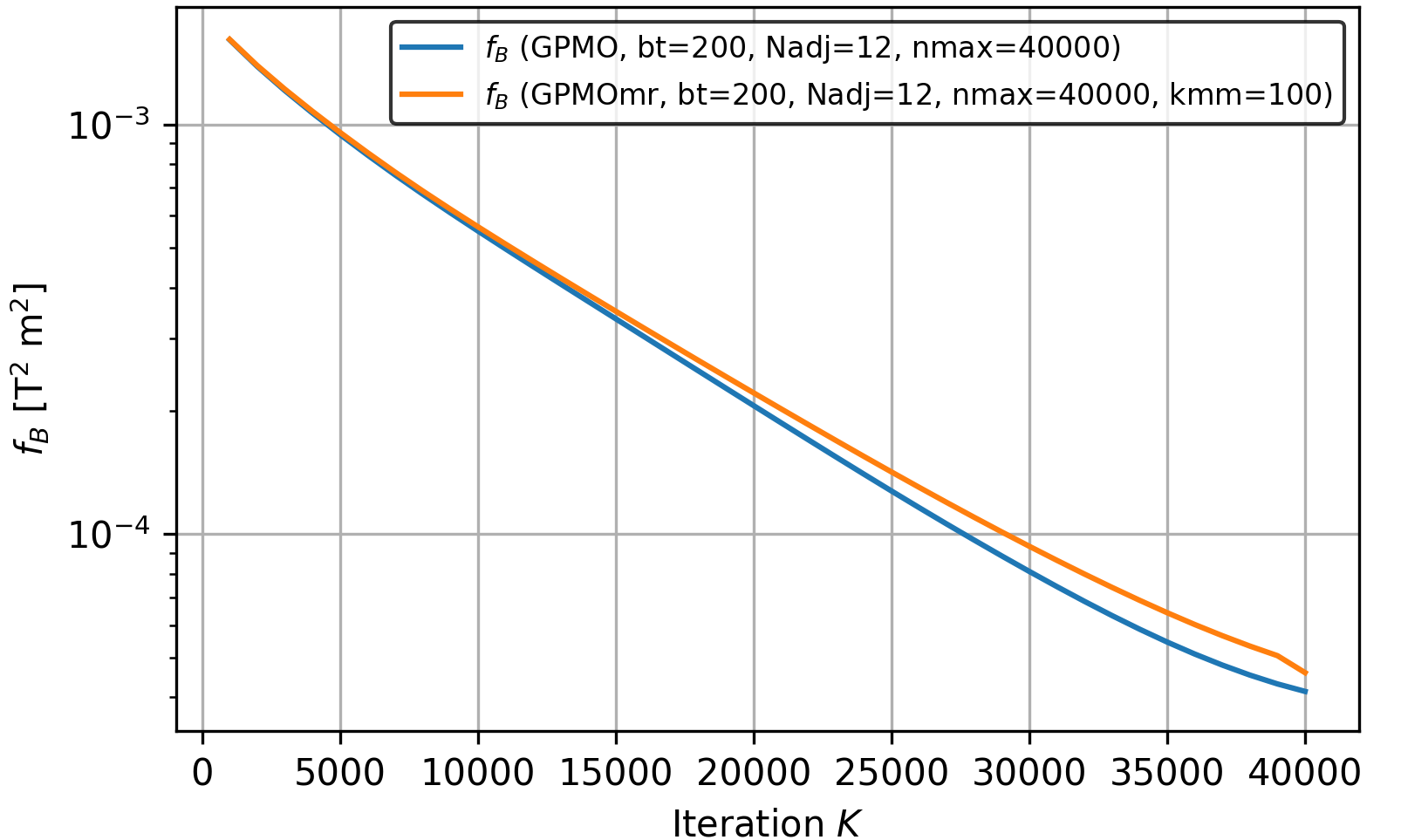}
    \caption{Squared-flux error $f_B$ versus greedy iteration $K$ for the GB50UH substitution study at $B_0\approx 0.5$~T. Classical (uncoupled) GPMO and GPMOmr converge on nearly identical trajectories, with GPMOmr saturating at a slightly higher final error, consistent with earlier MUSE comparisons.}
    \label{fig:gb50uh-mse-history}
\end{figure}

The squared-flux histories for the GB50UH runs are shown in Fig.~\ref{fig:gb50uh-mse-history}. Both algorithms converge smoothly on the same logarithmic trend. As in the previous MUSE comparisons, the macromagnetic-refinement run converges to a slightly higher final error than the classical rigid-remanence run. At the end of the $K=5\times 10^4$ runs we obtain
\begin{equation}
\begin{aligned}
f_B^{\mathrm{GPMO}} &\approx 4.1321\times 10^{-5}, \\
f_B^{\mathrm{GPMOmr}} &\approx 4.5916\times 10^{-5}.
\end{aligned}
\end{equation}
so the fully coupled macromagnetic loop remains within about $10\%$ of classical GPMO while operating in a significantly stronger applied-field environment. The two curves track one another closely over the full run, and the small separation that persists at late iterations is consistent with the behavior observed in earlier sections.


A pointwise comparison of the final dipole-moment fields shows the same qualitative structure seen in the baseline MUSE studies. The histogram of $\Delta M_i$ again separates into four buckets associated with (i) sites that agree nearly exactly, (ii) sites active in only one run ($\Delta M\sim M_{\max}$), (iii) approximately orthogonal placements ($\Delta M\sim \sqrt{2}\,M_{\max}$), and (iv) rare anti-parallel placements ($\Delta M\sim 2\,M_{\max}$). For GB50UH the near-zero bucket is substantially more dominant: among sites active in at least one run, $95.45\%$ fall near $\Delta M\approx 0$, while $3.47\%$ lie near $M_{\max}$, $1.08\%$ lie near $\sqrt{2}\,M_{\max}$, and only a single site is consistent with the $2M_{\max}$ bucket. This shift toward the near-zero peak indicates that, in this long-run setting with a less sparse final configuration, classical GPMO and GPMOmr select more similar placements and orientations, and the largest GPMOmr-driven redistributions are reduced compared with the sparser baseline runs.

\subsection{Demagnetization during operation}
\label{sec:results-alnico}

A second motivation for this work is to test the coupled macromagnetic loop in a regime where the rigid remanence assumption is least reliable, namely when demagnetization can meaningfully alter the effective magnetization during operation. As a representative low remanence alternative to NdFeB, we therefore repeat the MUSE optimization using an AlNiCo grade. While AlNiCo magnets have substantially lower remanence than modern NdFeB, some grades, notably Alnico 8HC, have a relatively large resistance to demagnetization within the AlNiCo family, with characteristic demagnetizing fields on the order of $10^{-1}$~T, suggesting they may plausibly tolerate modest applied fields from the MUSE coils when operated conservatively~\cite{BuntingAlnico2016}.

Concretely, we use a representative AlNiCo datasheet grade with remanence $B_r \approx 0.72$~T (Alnico 8HC, ACAT36J)~\cite{BuntingAlnico2016}, and set $B_{\max}=0.72$~T. As in the GB50UH case, the MUSE candidate grid stores reference dipole moment scales for the baseline material, so we rescale the maximum per site dipole moment in proportion to $B_{\max}$ while keeping the cell volume fixed. Using the same notation as above,
\begin{equation}
m_{\max}^{(\mathrm{AlNiCo})}
=
m_{\max}^{(\mathrm{base})}\,
\frac{B_{\max}^{(\mathrm{AlNiCo})}}{B_{\max}^{(\mathrm{base})}}.
\end{equation}
For reference, taking $m_{\max}^{(\mathrm{base})}\approx 7.5\times 10^{-2}$~A$\cdot$m$^{2}$ at $B_{\max}^{(\mathrm{base})}=1.465$~T gives
\begin{equation}
m_{\max}^{(\mathrm{AlNiCo})}
\approx
(7.5\times 10^{-2})
\frac{0.72}{1.465}
\simeq 3.7\times 10^{-2}\ \mathrm{A\cdot m^2},
\end{equation}
which sets the characteristic single dipole scale that appears in the difference histograms below.

To limit demagnetization risk in this low remanence setting, the coil currents are scaled so that the average field on the major radius is $B_0\approx 0.05$~T. In the macromagnetic refinement loop we use an isotropic permeability choice $\mu_{\mathrm{ea}}=\mu_{\mathrm{oa}}=3$, which increases the strength of magnet magnet coupling and demagnetizing response relative to the NdFeB runs. Both coil coupling and demagnetizing interactions are enabled for GPMOmr, while classical GPMO remains a rigid remanence optimization.

The AlNiCo runs use $n_\phi=64$ with $K=2.5\times 10^4$ greedy iterations. Backtracking is performed every 200 iterations with $N_{\mathrm{adjacent}}=12$ and a limit of $4\times 10^4$ nonzero magnets. Macromagnetic refinement is performed every $k_{\mathrm{mm}}=50$ iterations.

\begin{figure*}[t]
    \centering
    \includegraphics[width=\textwidth,height=0.34\textheight,keepaspectratio]{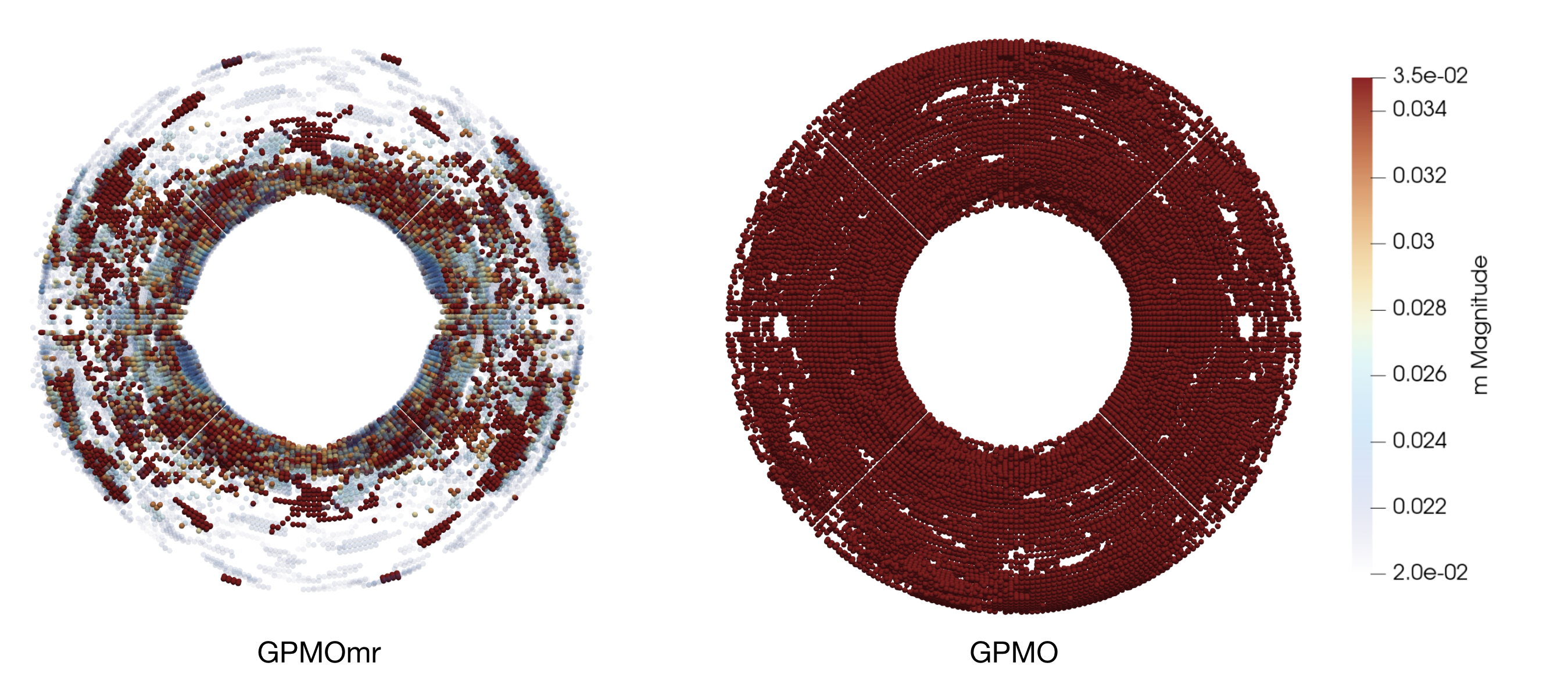}
    \caption{Glyph plots of the final magnetization patterns on the MUSE PM grid for the AlNiCo substitution study ($B_{\max}=0.72$~T). Colors indicate the dipole-moment magnitude $m$. The macromagnetic refinement run uses $\mu_{\mathrm{ea}}=\mu_{\mathrm{oa}}=3$ and exhibits stronger demagnetization relative to the rigid remanence GPMO solution.}
    \label{fig:alnico-pm-grid}
\end{figure*}

Figure~\ref{fig:alnico-pm-grid} shows the final magnetization patterns. The classical GPMO solution drives the dipole magnitudes rapidly toward their maximum allowed values, producing a nearly saturated pattern. In contrast, the macromagnetic refinement solution exhibits visibly stronger demagnetization: many tiles settle to reduced effective magnetization magnitudes under the coupled equilibrium solve. This behavior is consistent with the larger permeability used here, which amplifies the sensitivity of the equilibrium magnetization to local demagnetizing and applied fields, and it also makes backtracking more consequential, since partially demagnetized placements can alter which nearby dipoles are judged redundant or conflicting.

\begin{figure}[H]
    \centering
    \includegraphics[width=\columnwidth]{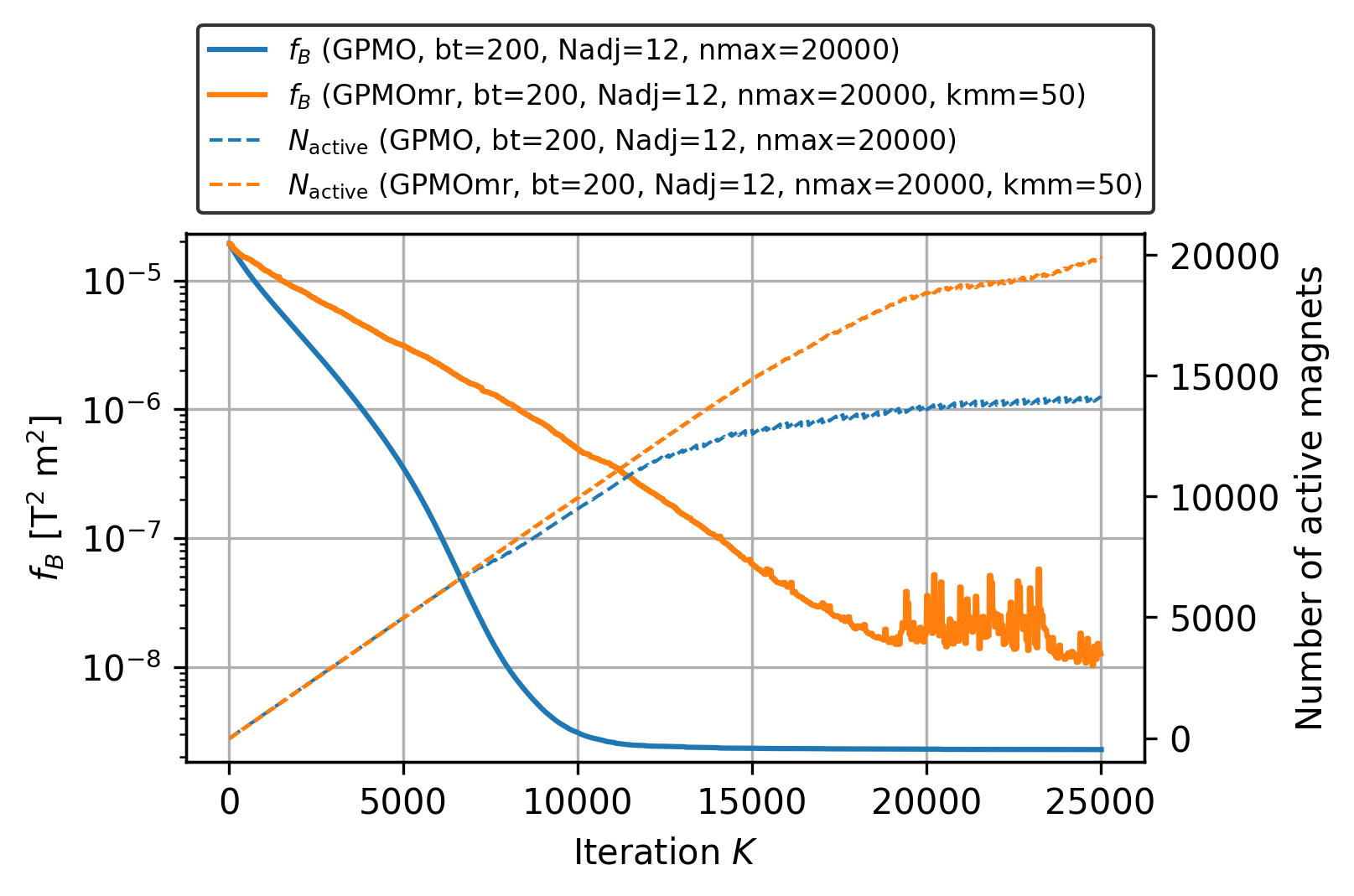}
    \caption{Squared-flux error $f_B$ versus greedy iteration $K$ for the AlNiCo substitution study at $B_0\approx 0.05$~T. The upper axis reports $N_{\mathrm{active}}$, the number of magnets placed (active dipoles) at each iteration. Classical GPMO decreases rapidly to a lower final error, while GPMOmr saturates higher and exhibits increased late-iteration variability, consistent with stronger coupling and demagnetization response for $\mu_{\mathrm{ea}}=\mu_{\mathrm{oa}}=3$.}
    \label{fig:alnico-mse-history}
\end{figure}

The squared flux histories are shown in Fig.~\ref{fig:alnico-mse-history}. Both runs decrease steadily, but the classical rigid remanence GPMO curve drops more quickly and reaches a lower final error, while the macromagnetic refinement curve saturates at a higher level and exhibits increased late iteration variability. Notably, once the GPMOmr run reaches $K\approx 1.8\times 10^4$, its $f_B$ begins to show pronounced spikes, indicating that late greedy additions are overdriving the coupled equilibrium and intermittently undoing surface error reductions after refinement and backtracking. This is also why we evaluate $\Delta M$ at $K=1.8\times 10^4$ below, as a representative snapshot just before this late run instability becomes dominant. In terms of the number of magnets placed, the classical GPMO curve is effectively converged by around $N_{\mathrm{active}}\approx 1.5\times 10^4$, while pushing toward $N_{\mathrm{active}}\approx 1.9\times 10^4$ yields no meaningful further improvement in $f_B$.

At the end of the $K=2.5\times 10^4$ iterations we obtain
\begin{equation}
\begin{aligned}
f_B^{\mathrm{GPMO}} &\approx 2.2793\times 10^{-9}, \\
f_B^{\mathrm{GPMOmr}} &\approx 1.2608\times 10^{-8},
\end{aligned}
\end{equation}
so the coupled macromagnetic loop ends at an error that is $5.5$ times larger than the idealized rigid remanence solution. This gap is expected: classical GPMO can exploit fully saturated dipoles everywhere, whereas in GPMOmr the coupled equilibrium solve reduces the effective magnetization at a subset of tiles, lowering the achievable net cancellation of $\mathbf B\cdot\mathbf n$.

\begin{figure*}[t] 
\centering \includegraphics[width=\textwidth,height=0.30\textheight,keepaspectratio]{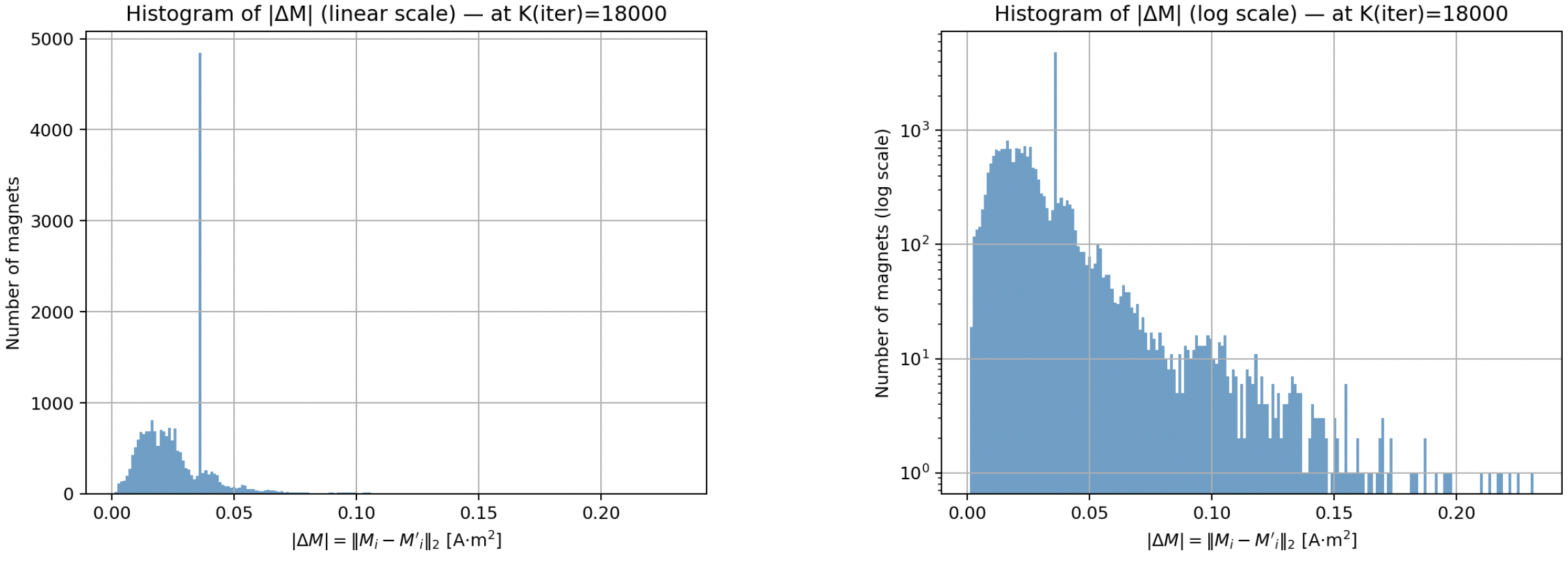} \caption{Histograms of the pointwise dipole-moment difference $\Delta M_i$ between GPMOmr and classical GPMO for the AlNiCo substitution study (linear and log $y$ scales), evaluated at $K=1.8\times 10^4$. The pronounced spike near $\Delta M \approx m_{\max}^{(\mathrm{AlNiCo})}\simeq 3.7\times 10^{-2}$~A$\cdot$m$^{2}$ indicates sites where one run places a nearly saturated dipole while the other leaves the site effectively empty (or strongly demagnetized); the peak height implies on the order of $5\times 10^{3}$ such site-level disagreements in this snapshot. Compared with the NdFeB cases, the remainder of the distribution is substantially broadened, reflecting widespread, continuous reductions in effective dipole magnitudes under the stronger macromagnetic response rather than sharply separated buckets.} \label{fig:alnico-deltaM} 
\end{figure*}

To quantify how the coupled equilibrium reshapes the final layout and amplitudes, we again compute the pointwise difference
\begin{equation}
\Delta M_i=\|\mathbf M_i^{\mathrm{GPMOmr}}-\mathbf M_i^{\mathrm{GPMO}}\|_2,
\end{equation}
where $\mathbf M_i$ denotes the dipole-moment vector at grid site $i$ (in A$\cdot$m$^{2}$). Figure~\ref{fig:alnico-deltaM} shows histograms of $\Delta M_i$ on both linear and logarithmic $y$ scales, evaluated at $K=1.8\times 10^4$ to capture the distribution before the late iteration $f_B$ spikes in Fig.~\ref{fig:alnico-mse-history} become dominant.

The spike near $\Delta M\approx M_{\max}^{(\mathrm{AlNiCo})}$ provides a direct diagnostic in this low remanence regime. Since a site that differs primarily by an on off decision contributes $\Delta M \approx M_{\max}^{(\mathrm{AlNiCo})}$, the height of this peak implies that roughly five thousand grid locations are magnetized in one solution but not in the other at this stage of the run. At the same time, the overall broadening of $\Delta M$ is consistent with Fig.~\ref{fig:alnico-pm-grid}: in the macromagnetic refinement solution, many tiles no longer cluster near a single saturation level but instead spread across a continuum of reduced effective moments. Thus, in the AlNiCo setting the discrepancy between classical GPMO and GPMOmr is driven both by site level selection differences and by widespread demagnetization driven amplitude redistribution, which together explain the larger separation in achievable $f_B$ observed in Fig.~\ref{fig:alnico-mse-history}.

\paragraph*
\noindent As an additional check in this low-remanence regime, we postprocess the \emph{fixed} AlNiCo layout with the standalone macromagnetic postprocessing script, recomputing the equilibrium magnetizations for the same set of active dipoles under the three coupling assumptions (unc, mm, mc) used throughout this paper. The evaluation uses the same coarse surface resolution as the optimization loop, $n_\phi=n_\theta=64$, and the same coil-current scaling ($B_0\simeq 0.05$~T on the major radius), so that changes in $\mathbf B\cdot\mathbf n$ and in $f_B$ can be interpreted directly as the effect of enforcing macromagnetic self-consistency on an otherwise identical layout.

\begin{figure}[t]
    \centering
    \includegraphics[width=\columnwidth]{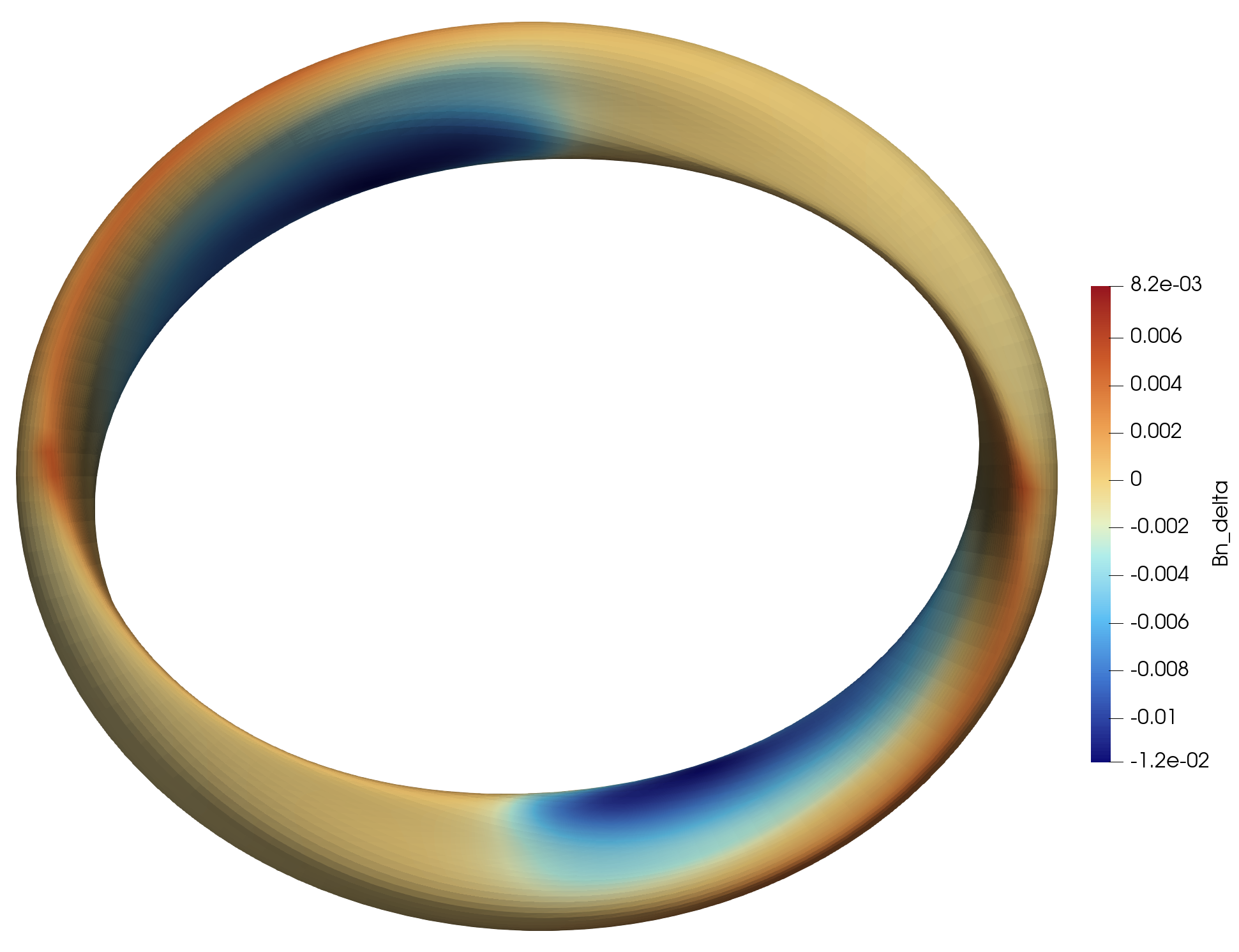}
    \caption{AlNiCo postprocessing diagnostic: $\Delta(\mathbf B\cdot\mathbf n)=(\mathbf B\cdot\mathbf n)_{\mathrm{unc}}-(\mathbf B\cdot\mathbf n)_{\mathrm{mc}}$ on the target surface for the fixed AlNiCo layout, with $B_0\simeq 0.05$~T and $n_\phi=n_\theta=64$.}
    \label{fig:alnico-post-bn-delta}
\end{figure}

Figure~\ref{fig:alnico-post-bn-delta} shows the corresponding change in surface normal field between the rigid-remanence and fully coupled postprocessed fields,
$\Delta(\mathbf B\cdot\mathbf n) = (\mathbf B\cdot\mathbf n)_{\mathrm{unc}}-(\mathbf B\cdot\mathbf n)_{\mathrm{mc}}$.
The extrema on this plot are $\max \Delta(\mathbf B\cdot\mathbf n)\approx 8.2\times 10^{-3}$~T and
$\min \Delta(\mathbf B\cdot\mathbf n)\approx -1.2\times 10^{-2}$~T, corresponding to a peak normalized change
\[
\max\frac{|\Delta(\mathbf B\cdot\mathbf n)|}{B_0}
\approx \frac{1.2\times 10^{-2}}{5.0\times 10^{-2}}
\simeq 2.37\times 10^{-1}.
\]
In other words, for AlNiCo with $\mu_{\mathrm{ea}}=\mu_{\mathrm{oa}}=3$, enforcing magnet--magnet and coil coupling produces \emph{tens-of-percent} changes in the local normal field relative to the coil-scale $B_0$, far larger than the percent-level perturbations seen in the NdFeB postprocessing.

More importantly, the global squared-flux objective deteriorates by several orders of magnitude when the same layout is re-equilibrated under macromagnetic coupling:
\begin{align}
f_B^{\mathrm{unc}} &= 2.280748\times 10^{-9}, \nonumber\\
f_B^{\mathrm{mm}}  &= 5.568253\times 10^{-6}, \nonumber\\
f_B^{\mathrm{mc}}  &= 6.606339\times 10^{-6}. \label{eq:alnico_postprocess_fB}
\end{align}
Equivalently, $f_B^{\mathrm{mm}}/f_B^{\mathrm{unc}}\approx 2.44\times 10^{3}$ and
$f_B^{\mathrm{mc}}/f_B^{\mathrm{unc}}\approx 2.90\times 10^{3}$, i.e.\ a degradation of $\approx 3.4$ orders of magnitude. This is the key failure mode in the AlNiCo setting: a layout that is ``excellent'' under the rigid-remanence model (unc) becomes \emph{orders-of-magnitude worse} once the same magnets are required to satisfy the coupled macromagnetic equilibrium (mm/mc). This observation is consistent with the behavior in Fig.~\ref{fig:alnico-mse-history}: late greedy additions can substantially re-equilibrate nearby moments under coupling, intermittently undoing surface-error reductions and leading to the observed $f_B$ spikes.

\FloatBarrier

\section{Conclusion}

Permanent magnets offer a promising route to simplified stellarator coils, but most design pipelines still rely on an idealized rigid-remanence approximation. In this work we have constructed and analyzed a device-scale macromagnetic model that relaxes this approximation by incorporating finite-permeability and demagnetizing interactions at the block level. Starting from a thermodynamically consistent work functional, we derived a quadratic susceptibility penalty, assembled the corresponding free-energy functional, and showed that macromagnetic equilibrium reduces to a single linear system $A\mathbf M = b$ for the block magnetizations. This formulation retains the essential physics of anisotropic hard magnets while remaining tractable for large arrays.

Applied in postprocessing to the published MUSE PM grid, the macromagnetic model produces small but measurable corrections. Individual bricks experience degree-scale tilts and $1$--$3\%$ changes in magnitude relative to their remanent state, with magnitude changes providing the dominant channel. These local adjustments map into $\mathcal{O}(10^{-3})$~T changes in the surface-normal field $\mathbf B\cdot\mathbf n$, corresponding to pointwise differences at the $\sim1\%$ level relative to the $0.15$~T target and rms changes of only a few percent. The resulting $\Delta(\mathbf B\cdot\mathbf n)$ pattern is smooth and does not introduce new localized defects, in line with earlier MagTense-based assessments that finite-$\mu$ effects on global metrics are modest and can be compensated by small coil retunings~\cite{Bjoerk2021,Qian2023}. At the same time, for a fixed GPMO layout the integrated squared-flux objective $f_B$ increases by more than a factor of two when macromagnetics is included, highlighting that even percent-level pointwise changes can matter once they are squared and integrated in the design metric.

Embedding the same macromagnetic model directly into the greedy permanent-magnet optimization algorithm through a winner-only refinement step yields GPMOmr layouts whose squared-flux errors $f_B$ remain within a few percent of the classical rigid-remanence solutions, both with and without backtracking and across refinement intervals $k_{\rm mm}\in\{1,25,50\}$. The $f_B(K)$ histories are nearly indistinguishable on a logarithmic scale, indicating that finite-permeability feedback does not significantly accelerate or degrade convergence for MUSE-like parameters. Nevertheless, the optimized arrays display visibly nonuniform block magnitudes and small tilts at tower fringes, and histogram diagnostics show that a nontrivial subset of sites is used in qualitatively different ways by GPMO and GPMOmr. In this sense the macromagnetic refinement redistributes demagnetizing stresses internally while preserving essentially the same plasma-facing performance.

Beyond the baseline MUSE comparisons, the two material and field-strength examples illustrate how the coupled model can be used as an engineering-facing stress test rather than a purely diagnostic correction. All quoted comparisons in this paragraph (including the final $f_B$ values and any ``unc/mm/mc'' deltas) are taken from the final post-processing pipeline, using the same evaluation conventions as in the baseline MUSE case (in particular, consistent current rescaling to the target $B_0$ and a fixed surface-error evaluation resolution). In the higher-field GB50UH example ($B_0\approx 0.5$~T), macromagnetic refinement remains close to rigid remanence in plasma-facing metrics (within $\sim10\%$ in the final $f_B$ for the configurations considered) and does not show any device-scale collapse of dipole magnitudes, consistent with the larger demagnetization margin of this grade at room temperature~\cite{Arnold2021GBDNeoCatalog}. By contrast, in the AlNiCo example ($B_0\approx 0.05$~T with a deliberately stronger effective permeability model), macromagnetic refinement produces widespread reductions in effective dipole magnitudes and a substantially larger gap relative to the idealized rigid-remanence solution, highlighting that coupling and demagnetization become central when the material response is less ``hard'' or when the macromagnetic feedback is amplified. In this sense, the coupled solver is not merely ``correcting'' a design, but actively identifying when a seemingly excellent rigid-remanence layout is fragile under a more realistic material response.

These findings support two main conclusions. First, for MUSE-like hard-magnet parameters, macromagnetic corrections to surface $\mathbf B\cdot\mathbf n$ are small yet quantitatively important at the percent level in global metrics and can be captured efficiently by a block-level linear model. Second, incorporating those corrections into the synthesis loop via GPMOmr yields designs that are very similar to classical GPMO in terms of global error measures, but that differ in how and where magnet strength is deployed across the grid. This suggests a useful division of labor: classical GPMO provides a fast first pass to identify good discrete layouts, while macromagnetics refines how those layouts are populated and can be used to quantify robustness margins or to enforce additional engineering criteria.

The present work also raises several directions for further study. On the engineering side, our results suggest value in increasing the effective degrees of freedom of the PM array, for example through multi-orientation holders, segmented towers, or graded-remanence materials that allow macromagnetics to redistribute field more flexibly. Another natural direction is to move away from full toroidal grids and instead form localized \emph{permanent-magnet islands} in regions where the coil field is already strong, using macromagnetics to exploit these higher-field working points more aggressively and to reduce the total number of bricks. This connects naturally to sensitivity-based tools that diagnose and control \emph{plasma magnetic islands} on the target surface~\cite{Chambliss2025}, and could guide the placement and shaping of permanent-magnet islands around coils.

Finally, our modeling assumptions are explicitly tailored to hard magnets: strong uniaxial anisotropy locks the easy axis to the lattice, susceptibilities remain modest, and the response is well approximated by a reversible, linear tensor $\boldsymbol\chi$ at the block scale. In soft or partially soft materials, by contrast, anisotropy is weaker, susceptibilities can be large and strongly field dependent, and domain-wall motion and hysteresis become central. In that regime the linear system $A\mathbf M = b$ ceases to be an adequate description; the equilibrium becomes nonlinear, history dependent, and more tightly coupled to micromagnetic length scales. Extending the present framework to such materials will likely require constitutive relations based on nonlinear $B(H)$ curves that capture hysteresis and domain-wall motion, together with a more explicit coupling to micromagnetic physics. We expect that such extensions, together with applications to higher-field permanent-magnet stellarators, hybrid coil and PM configurations, and coil-centered permanent-magnet islands, will help further close the gap between idealized design and experimentally realized fields.

\begin{acknowledgments}
This work was supported by the Simons Foundation under award 560651.
The authors gratefully acknowledge code and assistance from Tony Qian and 
Xu Chu. The authors thank the MUSE team for making their stellarator configurations and permanent-magnet designs available, and the developers of SIMSOPT and related software for providing an open computational ecosystem for stellarator optimization. Thanks to the MagTense team and 
Andrea Roberto Insinga for suggestions. Thanks to the Courant Institute's SURE program for partially funding this work.
\end{acknowledgments}

\appendix
\section{Reversible Magnetic-Work Functional: first-principles derivation}
\label{app:reversible-work}
\addcontentsline{toc}{section}{Appendix A: Derivation of the Reversible Magnetic-Work Functional}

We derive the reversible magnetic-work functional directly from Maxwell's equations and Poynting's theorem, following the standard continuum electrodynamics convention~\cite{Jackson,LandauECM,Stratton}. Throughout we assume quasi-static, isothermal, and reversible (non-hysteretic) processes: fields vary slowly in time, displacement currents and eddy currents are negligible, temperature is fixed ($dT=0$), and the material response is reversible so that the work is path-independent~\cite[Sec.~6.7]{Jackson}\cite[Secs.~15--16]{LandauECM}.

The instantaneous mechanical power supplied by external current sources is
\begin{equation}
P(t)=\int_V \mathbf J_{\rm ext}\!\cdot\!\mathbf E\,dV 
\label{eq:A2}
\end{equation}
with $\mathbf J_{\rm ext}$ the source conduction current density. In the quasi-static limit $\partial\mathbf D/\partial t\simeq 0$, Amp\`ere's law reduces to
\begin{equation}
\mathbf J_{\rm ext}=\nabla\times\mathbf H 
\label{eq:A3}
\end{equation}
Substituting into Eq.~\eqref{eq:A2} gives
\begin{equation}
P(t)=\int_V (\nabla\times\mathbf H)\!\cdot\!\mathbf E\,dV 
\label{eq:A4}
\end{equation}
Applying the vector identity $\mathbf A\cdot(\nabla\times\mathbf B)=-\mathbf B\cdot(\nabla\times\mathbf A)+\nabla\cdot(\mathbf B\times\mathbf A)$ with $(\mathbf A,\mathbf B)=(\mathbf E,\mathbf H)$ yields
\begin{equation}
P(t)=-\int_V \mathbf H\cdot(\nabla\times\mathbf E)\,dV
     +\int_V \nabla\cdot(\mathbf E\times\mathbf H)\,dV .
\label{eq:A5}
\end{equation}
The second term is the outward Poynting flux. For a sufficiently large control surface in free space, or a perfectly conducting enclosure, this flux vanishes,
\begin{equation}
\int_{V}\nabla\!\cdot(\mathbf E\times\mathbf H)\,dV
=\oint_{\partial V} (\mathbf E\times\mathbf H)\!\cdot d\mathbf a=0 ,
\label{eq:A6}
\end{equation}
leaving
\begin{equation}
P(t)=-\int_V \mathbf H\cdot(\nabla\times\mathbf E)\,dV .
\label{eq:A7}
\end{equation}
Faraday's law in the quasi-static limit is
\begin{equation}
\nabla\times\mathbf E=-\frac{\partial\mathbf B}{\partial t} ,
\label{eq:A8}
\end{equation}
so Eq.~\eqref{eq:A7} becomes
\begin{equation}
P(t)=\int_V \mathbf H\cdot\frac{\partial\mathbf B}{\partial t}\,dV .
\label{eq:A9}
\end{equation}
Over an interval $dt$, the reversible work is $\delta W_{\rm rev}=P(t)\,dt$, and with $\delta\mathbf B=(\partial\mathbf B/\partial t)\,dt$ we obtain
\begin{equation}
\delta W_{\mathrm{rev}}
   =\int_V \mathbf H\cdot\delta\mathbf B\,dV ,
\label{eq:A1}
\end{equation}
which depends only on instantaneous fields and is path-independent under the assumed conditions~\cite[Sec.~15]{LandauECM}.

At constant temperature, the Helmholtz free-energy differential equals the reversible work,
\begin{equation}
dF=\delta W_{\rm rev},\qquad F=\int_V F_{\rm int}\,dV ,
\label{eq:A11-12}
\end{equation}
so that
\begin{equation}
dF=\int_V dF_{\rm int}\,dV .
\end{equation}
Using $\mathbf B=\mu_0(\mathbf H+\mathbf M)$ and $\delta\mathbf B=\mu_0(\delta\mathbf H+\delta\mathbf M)$ in Eq.~\eqref{eq:A1} gives
\begin{equation}
\delta W_{\rm rev}
=\mu_0\!\int_V \mathbf H\cdot\delta\mathbf H\,dV
+\mu_0\!\int_V \mathbf H\cdot\delta\mathbf M\,dV .
\label{eq:A14}
\end{equation}

Since $\mathbf H_a\cdot\delta\mathbf H_a=\tfrac12\delta(H_a^2)$, the first term is a reference contribution that depends only on the prescribed external sources (coil currents) and not on variations in $\mathbf M$,
\begin{equation}
\mu_0\!\int_V \mathbf H_a\cdot\delta\mathbf H_a\,dV
=\delta\!\left[\tfrac{\mu_0}{2}\int_V H_a^2\,dV\right] .
\end{equation}
Dropping this reference term leaves the internal free-energy density differential
\begin{equation}
dF_{\rm int}=\mu_0\,\mathbf H_a\cdot d\mathbf M .
\label{eq:A15}
\end{equation}
This is the SI-unit form used to build the susceptibility term in the macromagnetic functional (Sec.~III)~\cite[Secs.~15--16]{LandauECM}\cite[Ch.~2]{Jiles2016}.


An equivalent splitting can be written in terms of the ``vacuum'' field $\mathbf H_0$ that would exist in the absence of the material, with $\mathbf B_0=\mu_0\mathbf H_0$. One finds
\begin{equation}
\delta W_{\rm rev}
=\mu_0\!\int_V \mathbf H_0\cdot\delta\mathbf H_0\,dV
+\mu_0\!\int_V \mathbf H_0\cdot\delta\mathbf M\,dV ,
\end{equation}
so that, again discarding the $M$-independent reference term,
\begin{equation}
dF_{\rm int}=\mu_0\,\mathbf H_0\cdot d\mathbf M .
\end{equation}

\bigskip
\section{Uniaxial Susceptibility Quadratic Form: component expansion}
\label{app:quadratic-form}
\addcontentsline{toc}{section}{Appendix B: Component Expansion of the Susceptibility Quadratic Form}

In the macromagnetic model the local contribution of a single cell to the free-energy functional $F[\mathbf M]$ includes a susceptibility term of the form
\begin{equation}
F_{\chi}[\mathbf M]
= \tfrac{\mu_0}{2}\,(\mathbf M-\mathbf M_{\rm rem})\cdot\boldsymbol\chi^{-1}(\mathbf M-\mathbf M_{\rm rem}),
\end{equation}
so that $F[\mathbf M]$ contains a sum of such local contributions over all cells. In this appendix we analyze the structure of the quadratic form appearing in this susceptibility energy, and include the micromagnetic contributions as well, for completeness.

We expand the quadratic form
\begin{equation}
(\mathbf M-\mathbf M_{\rm rem})\cdot\boldsymbol\chi^{-1}(\mathbf M-\mathbf M_{\rm rem}),
\label{eq:B0}
\end{equation}
for a medium with a single easy axis $\hat{\mathbf u}$ and two independent susceptibilities: longitudinal $\chi_\parallel$ and transverse $\chi_\perp$~\cite[Ch.~6]{Jiles2016}.

Introduce an orthonormal triad $\{\hat{\mathbf e}_1,\hat{\mathbf e}_2,\hat{\mathbf u}\}$ with $\hat{\mathbf u}$ the easy axis. In this basis the inverse susceptibility tensor has diagonal form
\begin{equation}
\boldsymbol\chi^{-1}
=\begin{pmatrix}
\chi_\perp^{-1} & 0 & 0 \\
0 & \chi_\perp^{-1} & 0 \\
0 & 0 & \chi_\parallel^{-1}
\end{pmatrix}.
\label{eq:B1}
\end{equation}
The deviation from remanence is
\begin{equation}
\Delta\mathbf M=\mathbf M-\mathbf M_{\rm rem}
= M_1\hat{\mathbf e}_1+M_2\hat{\mathbf e}_2+(M_\parallel-M_{\rm rem})\hat{\mathbf u},
\label{eq:B2}
\end{equation}
where introduce the shorthand $\mathbf M_\perp=M_1\hat{\mathbf e}_1+M_2\hat{\mathbf e}_2$ for the transverse part and $\Delta M_\parallel=M_\parallel-M_{\rm rem}$ for the longitudinal deviation. Acting with Eq.~\eqref{eq:B1} gives 
\begin{equation}
\boldsymbol\chi^{-1}\Delta\mathbf M
=\chi_\perp^{-1}\mathbf M_\perp+\chi_\parallel^{-1}\Delta M_\parallel\,\hat{\mathbf u}.
\label{eq:B3}
\end{equation}
Taking the scalar product of Eq.~\eqref{eq:B2} with Eq.~\eqref{eq:B3} yields 
\begin{equation}
\Delta\mathbf M\cdot\boldsymbol\chi^{-1}\Delta\mathbf M
=\frac{\|\mathbf M_\perp\|^2}{\chi_\perp}
+\frac{(M_\parallel-M_{\rm rem})^2}{\chi_\parallel}.
\label{eq:B4}
\end{equation}
Substituting Eq.~\eqref{eq:B4} into the quadratic form \eqref{eq:B0} and multiplying by $\mu_0/2$ gives the susceptibility contribution to the free-energy density,
\begin{equation}
f_\chi(\mathbf M)
=\frac{\mu_0}{2}\left[\frac{\|\mathbf M_\perp\|^2}{\chi_\perp}
+\frac{(M_\parallel-M_{\rm rem})^2}{\chi_\parallel}\right] .
\label{eq:B5}
\end{equation}
Here $\mathbf H_a$ is the prescribed applied field from external sources (e.g.\ coils) and is held fixed under variations in $\mathbf M$, whereas the demagnetizing field $\mathbf H_d=\mathbf H_d[\mathbf M]$ is determined self-consistently by $\mathbf M$ through the magnetostatic equations.
Finally, combining exchange, demagnetizing, anisotropy, Zeeman, and susceptibility terms, the full free-energy functional is
\begin{multline}
F[\mathbf M]
= \int_\Omega \Biggl[
  \underbrace{\tfrac{A_{\rm ex}}{M_s^2}\|\nabla\mathbf M\|^2}_{\text{exchange}}
  - \underbrace{\tfrac{\mu_0}{2}\,\mathbf M\cdot\mathbf H_d}_{\text{demag}} \\[4pt]
  - \underbrace{K\tfrac{(\mathbf M\cdot\hat{\mathbf u})^2}{M_s^2}}_{\text{anisotropy}}
  - \underbrace{\mu_0\,\mathbf M\cdot\mathbf H_a}_{\text{Zeeman}} \\[4pt]
  + \underbrace{\tfrac{\mu_0}{2}
    (\mathbf M-\mathbf M_{\rm rem})
    \cdot \chi^{-1}(\mathbf M-\mathbf M_{\rm rem})}_{\text{susceptibility}}
\Biggr]\,dV
\label{eq:Fstar}
\end{multline}
Stationarity requires
\begin{multline}
\frac{\delta F}{\delta\mathbf M} = 0
\;\;\Longleftrightarrow\;\;
-\frac{2A_{\rm ex}}{M_s^2}\nabla^2\mathbf M
-\mu_0\,\mathbf H_d \\[4pt]
-\frac{2K}{M_s^2}(\mathbf M\cdot\hat{\mathbf u})\hat{\mathbf u}
-\mu_0\,\mathbf H_a
+\mu_0\,\chi^{-1}(\mathbf M-\mathbf M_{\rm rem}) = 0 ,
\label{eq:Heff_reduced}
\end{multline} 
or equivalently $\mathbf H_{\rm eff}\equiv-\mu_0^{-1}\,\delta F/\delta\mathbf M=0$.



\medskip
\medskip
\noindent\textit{Remark.}
The demagnetization contribution in Eq.~\eqref{eq:Fstar},
$-\tfrac{\mu_0}{2}\,\mathbf M\cdot\mathbf H_d$, may appear to be linear in the magnetization.
However, the demagnetizing field $\mathbf H_d$ is not an independent variable: it is determined
self-consistently by $\mathbf M$ through the magnetostatic equations
$\mathbf H_d=-\nabla\phi$ and $\nabla\!\cdot(\mathbf H_d+\mathbf M)=0$, together with appropriate
boundary conditions. Equivalently, $\mathbf H_d=\mathbf H_d[\mathbf M]$ is a linear functional of
$\mathbf M$. The associated energy
\[
E_d[\mathbf M]
= -\frac{\mu_0}{2}\int_\Omega \mathbf M\cdot\mathbf H_d[\mathbf M]\,dV
\]
is therefore a magnetostatic self-energy and is quadratic in $\mathbf M$.

The prefactor $1/2$ is fixed by variational consistency. Under a perturbation
$\mathbf M\mapsto \mathbf M+\varepsilon\,\delta\mathbf M$, linearity implies
$\delta\mathbf H_d=\mathbf H_d[\delta\mathbf M]$, so
\[
\delta E_d
= -\frac{\mu_0}{2}\int_\Omega
\Bigl(\delta\mathbf M\cdot\mathbf H_d[\mathbf M]
      + \mathbf M\cdot\mathbf H_d[\delta\mathbf M]\Bigr)\,dV .
\]
Using the reciprocity of the magnetostatic kernel,
\(
\int_\Omega \mathbf A\cdot\mathbf H_d[\mathbf B]\,dV
=
\int_\Omega \mathbf B\cdot\mathbf H_d[\mathbf A]\,dV,
\)
the two terms are equal, yielding
\[
\delta E_d
= -\mu_0\int_\Omega \delta\mathbf M\cdot\mathbf H_d[\mathbf M]\,dV,
\qquad
\frac{\delta E_d}{\delta\mathbf M}
= -\mu_0\,\mathbf H_d[\mathbf M].
\]
This is precisely the demagnetizing contribution appearing in the Euler--Lagrange equation
\eqref{eq:Heff_reduced}.

\FloatBarrier      
\section{Additional visualizations for surface \texorpdfstring{$\mathbf B\cdot\mathbf n$}{B·n} (no backtracking)}
\label{app:surface-bn-no-bt}
\addcontentsline{toc}{section}{Appendix C: Additional visualizations for surface $\mathbf B\cdot\mathbf n$ (no backtracking)}

\begin{figure}[!ht]
    \centering
    \begin{minipage}[c]{0.48\linewidth}
        \centering
        \includegraphics[width=\linewidth]{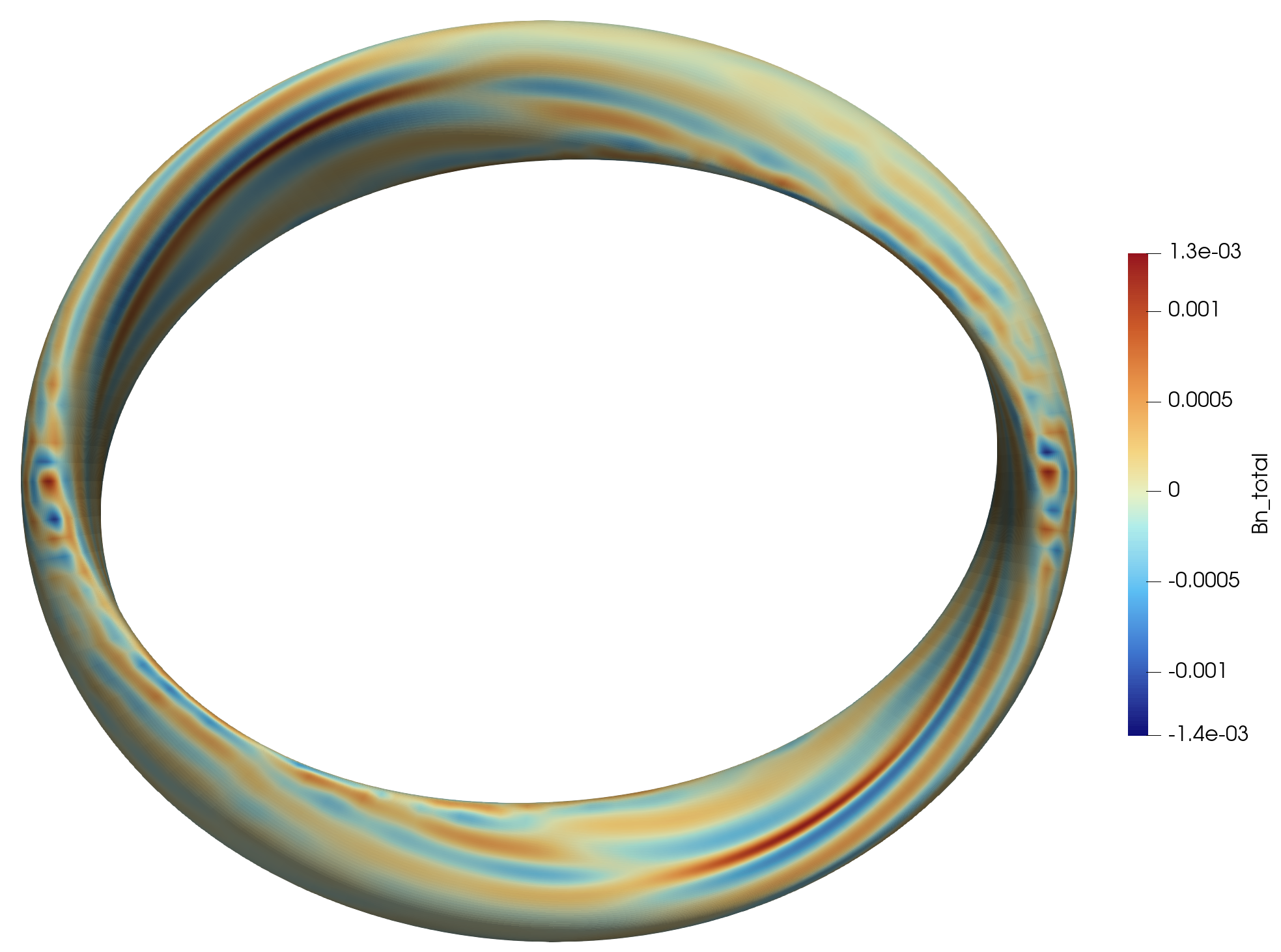}
    \end{minipage}
    \hfill
    \begin{minipage}[c]{0.48\linewidth}
        \centering
        \includegraphics[width=\linewidth]{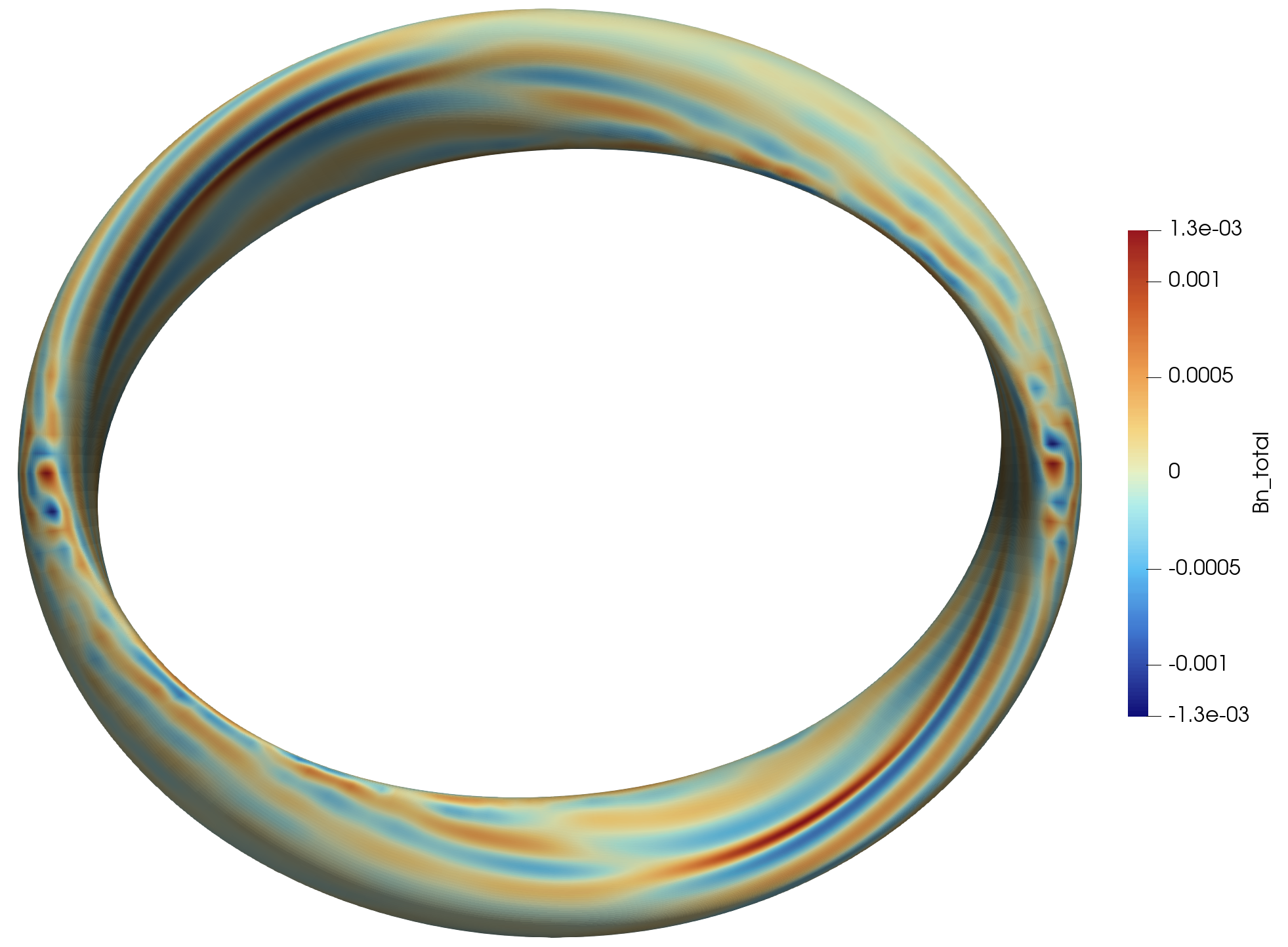}
    \end{minipage}

    \caption{Surface $\mathbf{B}\cdot\mathbf{n}$ residual for the classical (uncoupled) GPMO run (left) 
    and the GPMOmr run (right) on the MUSE PM grid without backtracking. The patterns closely match, 
    with only small local differences driven by macromagnetic corrections.}
    \label{fig:bn-gpmo-macro}
\end{figure}

For completeness we show here the visual diagnostics for the surface $\mathbf B\cdot\mathbf n$ residuals in the no-backtracking runs discussed in Sec.~\ref{sec:results}. The same plotting conventions and color scales as in Sec.~\ref{sec:results} are used throughout. Figure~\ref{fig:bn-gpmo-macro} shows the surface $\mathbf B\cdot\mathbf n$ residual for the classical (uncoupled) GPMO run and the GPMOmr run on the MUSE PM grid without backtracking.

\clearpage

\nocite{*}

\bibliography{refs}

\end{document}